\documentclass[aps,prb,preprint,showpacs]{revtex4-1}
\usepackage{graphicx}
\usepackage{bm}
\usepackage{amssymb}
\begin{document}
\title{Guided spin wave in monolayer CrSBr: Localization and spin-orbit coupling from dipolar field}
\author{D. Wang}
\email{daowei.wang@matfyz.cuni.cz}
\affiliation{Department of Condensed Matter Physics, Charles University, Ke Karlovu 5, Prague 2, 12116, Czech Republic}
\author{J. K. Vejpravov\'{a}}
\email{jana.vejpravova@matfyz.cuni.cz}
\affiliation{Department of Condensed Matter Physics, Charles University, Ke Karlovu 5, Prague 2, 12116, Czech Republic}
\date{\today}
\begin{abstract}
Spin-wave spectrum of monolayer CrSBr waveguides was studied by numerically diagonalizing the Bogoliubov-de Gennes Hamiltonian derived from linearising the Landau-Lifshitz-Gilbert equation. In contrast to its short-range counterparts, the long-range dipolar field acts statically as a confining potential for spin wave, while the dynamic part couples the spin and orbit degrees of freedom, thus giving rise to spin-orbit coupling for spin wave. Due to the inversion symmetry of the Hamiltonian and the spinor structure of the wave function, spin-wave eigenstates form doublets with definite parity. Micromagnetic simulation tallies well with numerical calculation. Our study on spin-wave eigenstates in CrSBr waveguides sheds light on the nature of exchange-dipole spin wave in ferromagnetic slabs. We confirm particularly that the robustness of the Damon-Eshbach mode is not derived from topology, but rather from the static dipolar field. Moreover, a thorough knowledge on spin wave in monolayer CrSBr itself represents a step forward to understanding the more complicated antiferromagnetic resonance in bulk CrSBr.
\end{abstract}
\maketitle

\section{Introduction}
\label{intro}
Due to their reduced dimensionality, two-dimensional (2D) materials can be affected more easily by external means such as gating electric field \cite{Novoselov04} and exhibit more quantum phenomena, one prominent example of which is the room-temperature quantum Hall effect \cite{Novoselov07}. In spintronics, this kind of versatility is long sought-for. Hence, the experimental discovery of a 2D magnet, CrI$_3$ \cite{Huang17}, led to intensive follow-up research in it and other 2D magnets. The catch of CrI$_3$ is its low Curie temperature, only about 45 K in the monolayer limit. With a relatively high N\'{e}el temperature of 132 K and air stability, CrSBr is a more promising van der Waals antiferromagnet for device integration \cite{Lee21,Ziebel24}.

Spin wave, or magnon after quantization, is magnetic elementary excitation that plays an important role in determining the magnetic behaviour of magnets. A well-known example is the Bloch $T^{3/2}$ law for the temperature dependence of magnetization at low temperature, caused by excitation of magnons. Spin wave is also pivotal in determining the low-frequency dynamics of magnetic topological textures. In particular, it was found that  inertial skyrmion behaviour can be induced by magnons through contact interaction with a skyrmion \cite{Wang22}. Moreover, it can be envisioned that dynamic interaction between skyrmions could be mediated by magnons acting as intermediate bosons, in analogy with electromagnetic interaction between electrons mediated by photons.

Electromagnetic field, mainly the magnetic part, is actually an important constituent of the total effective field acting on magnetization. This kind of magnetostatic, or dipolar, field is long-range and non-local in nature, in stark contrast to other constituents of the effective field, such as the anisotropy field and the Zeeman field, which are completely short-range and local. The exchange field can be approximated as a nearest-neighbour interaction, thus making it short-range, albeit non-local. The long-range nature of the dipolar field makes it difficult to be taken into account analytically. In the \textit{bona fide} 2D limit, a demagnetization factor $-1$ in the direction perpendicular to the 2D plane is enough to describe the dipolar field, contributing effectively as an easy-plane anisotropy. However, if the considered geometry is finite in any of the three dimensions, the dipolar field concentrates mostly around the boundary, due to magnetic charge accumulation therein. The localized dipolar field acts as a confining potential for spin wave and can cause spin-wave localization and effective spin-orbit coupling of magnons, similar to corresponding phenomena of electrons.

Terahertz spin wave in CrSBr was first measured by neutron scattering to extract the various exchange coupling constants \cite{Scheie22}. Due to the small interlayer antiferromagnetic (AFM) exchange coupling, AFM resonance in the gigahertz range can also be measured \cite{Cham22}. Recently, the effect of the dipolar field on the AFM resonance spectrum was considered \cite{Xu25csb}, but there is still a lack of a systematic treatment of the dipolar-field effect on spin wave in CrSBr. In view of its potential application as magnonic devices, we will hence consider here the exchange-dipole spin wave in monolayer CrSBr, highlighting the dipolar confinement effect on spin wave in monolayer and even multilayer \cite{Teuling25} devices with finite lateral extension. As bulk CrSBr can be viewed as monolayers coupled together through the AFM interlayer exchange interaction, our result on spin wave in monolayer CrSBr can shed light on the interpretation of the debated AFM resonance spectrum of bulk CrSBr \cite{Xu25csb}.

The organization of the article is as follows. After this Introduction, in Secs. \ref{para} and \ref{mgf}, we will give the magnetic parameters and the magnetostatic Green functions that will be used in our calculation and simulation. Sec. \ref{sw} will then present briefly the theoretical formulation of spin wave. Spin-wave spectra for field applied along principal crystallographic axes will be given in Secs.  \ref{swk}, \ref{swwg}, and  \ref{swslab} for an infinite film, a waveguide, and a slab of monolayer CrSBr, respectively. Our conclusion on the current investigation will be presented in Sec. \ref{conclusion}, which is preceded by a discussion about the physical origin of the Damon-Eshbach, or edge, mode in Sec. \ref{deorigin}.

\section{Magnetic parameters}
\label{para}
The space group of CrSBr is\textit{ Pmmn}, with orthorhombic lattice constants \cite{Yang21} $a = 3.54$ \AA, $b = 4.73$ {\AA}, and $c = 7.96$ \AA. The magnetic moment of CrSBr is mainly derived from Cr$^{3+}$ ions. Using the theoretically calculated value \cite{Yang21} 3 $\mu_B$/f.u., the saturation magnetization can be easily calculated as $M_s=$ 4.2 $\times$ 10$^5$ A/m, which is close to the experimentally measured bulk value \cite{Goser90}. The anisotropy constants along the $b$ and $c$ axes are $K_b$ = 5.3 $\times$ 10$^4$ J/m$^3$ and $K_c$ = 2.6 $\times$ 10$^4$ J/m$^3$, respectively, considering only the spin-orbit contribution to the anisotropy \cite{Yang21}. Our $K_b$ is slightly larger than the value 4.4 $\times$ 10$^4$ J/m$^3$ obtained from fitting the microwave absorption spectrum \cite{Cho23}. Taking into account of the demagnetization energy changes the $c$ axis into a hard axis, $K_c$ = $-$8.3  $\times$ 10$^4$ J/m$^3$, which is smaller than the value $-$9.7 $\times$ 10$^4$ J/m$^3$ determined experimentally \cite{Cho23}. Converting to anisotropy fields, $\mu_0 H_b$ = 0.25 T and $\mu_0 H_c$ = $-$0.40 T, which are significantly smaller than $\mu_0 H_b$ = 0.38 T and $\mu_0 H_c$ = $-$0.92 T obtained from AFM resonance measurements \cite{Cham22}. The exchange interaction between in-plane ($ab$) Cr$^{3+}$ ions is mediated through S$^{2-}$ or Br$^-$ ions or both. Between two monolayers stacked along the $c$ direction, the exchange interaction is mediated by two Br$^-$ ions through the exchange path Cr-Br-Br-Cr, making the interlayer exchange coupling much smaller \cite{Scheie22}. Using the fitted nearest-neighbour Heisenberg exchange energy $J=1.67$ meV and $J=1.90$ meV along $a$ and $b$ directions respectively,  the corresponding micromagnetic exchange constants can be calculated to be $A=1.26$ pJ/m and $A=1.95$ pJ/m, using the relation \cite{Exl21} $A = J S^2/d$ with $d$ equal to $a$ or $b$ and $S=3/2$. As the actual crystal symmetry is rhombic instead of cubic, the exchange interactions along $a$ and $b$ directions are different. This anisotropic character of the exchange interaction can give rise to anisotropic micromagnetic exchange interactions that will make Hopfions stable \cite{Rybakov22}. As the focus of the current study is on low-frequency spin wave, the complications and opportunities accompanying the anisotropic character of CrSBr will not be considered, and we approximate the anisotropic CrSBr as an isotropic Heisenberg magnet. The isotropic exchange constant is taken to be the algebraic average $A$ = 1.6 pJ/m.

\begin{figure}\centering
\begin{minipage}[c]{0.8\linewidth}
\includegraphics[width=\linewidth]{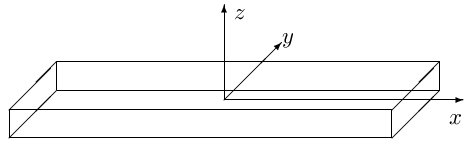}
\end{minipage}
\caption{Coordinate system used for spin wave in a monolayer of CrSBr. The rectangular slab is bounded by coordinate planes $x = \pm l$, $y = \pm w$, and $z = \pm t$. In the limit $l \rightarrow \infty$, the slab becomes an ideal, infinitely long, waveguide. An extended film has both $l \rightarrow \infty$ and $w$ $\rightarrow \infty$.}
\label{xyz}
\end{figure}

\section{Magnetostatic Green function}
\label{mgf}
The dipolar field $\textbf {d}$ normalized to $M_s$ depends on the magnetization configuration through the divergence-free Maxwell equation $\nabla \cdot (\textbf {m} - \textbf {d}) = 0$, where \textbf {m} is the unit magnetization vector, $\textbf {m}=\textbf {M}/M_s$. An explicit relation between $\textbf {d}$ and $\textbf {m}$ can be obtained using the magnetostatic Green function \cite{Guslienko11}, $d_i (\textbf {r}) = (G_{ij} *  m_j) (\textbf{r})$. $G _{ij}$ is a real symmetric tensor, where $i$ and $j$ can be any of the three subscripts $x$, $y$, and $z$. The star operator $*$ following $G_{ij}$ designates convolution in coordinate space, $(G _{ij} * m_j) (\textbf {r}) = \int d \textbf {r}' G _{ij} (\textbf{r} - \textbf{r}') m_j (\textbf{r}')$, where $\textbf{r}$ and $\textbf{r}'$ are general position vectors. The convolution operation demonstrates clearly the long-range characteristics of the dipolar field: $\textbf{d}$ at one point is determined by the whole distribution of $\textbf{m}$, rather than the local value or differentiation of $\textbf{m}$. Whenever there is no risk of confusion, we will omit the argument of $G _{ij}$. The dipolar field is given by the convolution of $G _{ij}$ and $\textbf{m}$ demonstrates one important feature of the magnetostatic interaction: The dipolar field is conformal, or scale, invariant; it depends only on the relative size rather than the absolute dimension of a sample. In addition, $G_{ij}$ satisfies the constraint $G _{ii} (\textbf{r}-\textbf{r}')=\delta (\textbf{r}-\textbf{r}')$, which reduces the independent components of $G_{ij}$ to five: $G _{xx}$ can be obtained from $G _{yy}$ and $G _{zz}$ as $G _{xx} = \delta (\textbf{r}-\textbf{r}') - G _{yy} - G _{zz}$. For the quasi-2D case of a thin slab of magnetic material (c.f. Fig. \ref{xyz}), $G_{ij}$'s dependence on $z-z'$ can be eliminated, through the simplifying assumption that the magnetization distribution in the thickness ($z$) direction is uniform. In this approximation, $G_{ij}$ depends only on the difference of the in-plane position vectors $\bm{\rho} = \hat {x} x + \hat{y} y$ and $\bm {\rho}' = \hat {x} x' + \hat{y} y'$. $\hat {x}$ and $\hat {y}$ are unit vectors along the $x$ (or crystallographic $a$) and $y$ (or crystallographic $b$) directions.

For the study of spin wave, it is more convenient to use the Fourier transform of $G_{ij}$, which is defined as $G_{ij}(\textbf{k}) = \int d \bm {\rho} G_{ij} (\bm {\rho}) \exp (-i \textbf{k} \cdot \bm{\rho})$. The advantage of using $\textbf{k}$ instead of $\bm{\rho}$ can be traced back to the convolution structure $(G_{ij} * m_j) (\bm{\rho})$ which will become product in \textbf {k}-space, $G _{ij} (\textbf{k}) m_j (\textbf{k})$, upon Fourier transformation. Explicitly, $G_{zz}(k) = 1 - f_k, G_{ij} (\textbf{k}) = k_i k _j f _k/ k^2$ where the function \cite{Guslienko11} $f _k = 1 - (1 - e^{- 2 k t}) /2 k t$ results from averaging over the thickness $2 t$ along $z$, and $k = \sqrt {k_x^2 + k_y^2}$ is the modulus of $\textbf{k}$. Meanwhile, the independent components of $G_{ij}$ are reduced further from five to three: Off-diagonal components $G_{zi}$ are averaged to zero. $G_{ij} (\textbf{k})$ are involved in description of spin wave in infinitely extended films.

\begin{figure}\centering
\begin{minipage}[c]{0.8\linewidth}
\includegraphics[width=\linewidth]{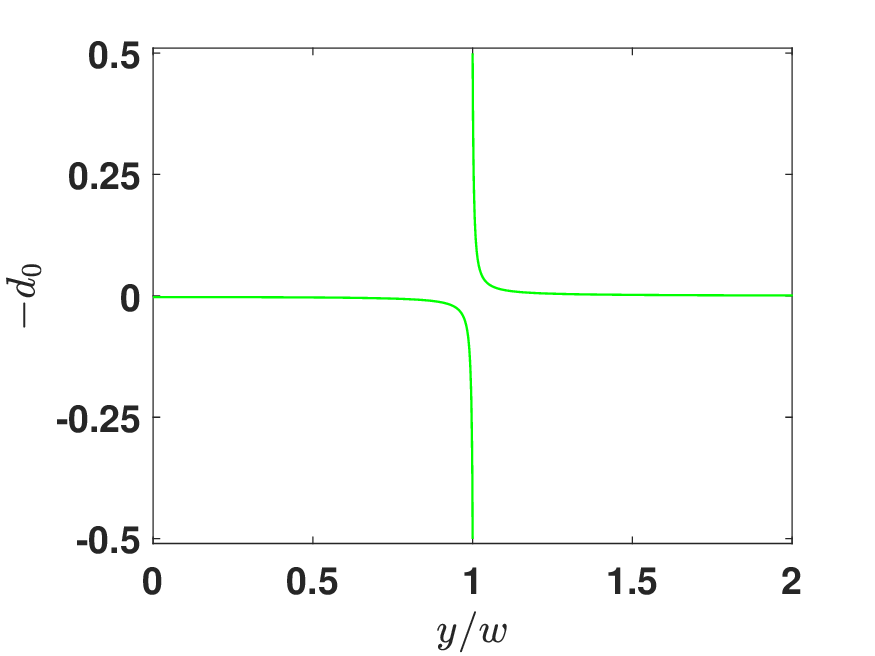}
\end{minipage}
\caption{Static dipolar field for a wave guide when $\textbf {m}  = \hat {y}$, Eq (\ref{yd0}). The jump at $y = w$ is exactly 1, as required by the continuity of the \textbf {B} field. $-d _0$ is symmetric with respect to $y$.}
\label{ydemag}
\end{figure}

For the configuration of a waveguide with width $2 w$ along $\hat {y}$, which will be the focus of the current study, we need the $x$-Fourier-transformed components $G_{ij}(k_x,y) = \int dx G_{ij} (\bm {\rho}) \exp (-i x k_x)$. The $zz$ component can be calculated analytically \cite{Guslienko11}, $G_{zz} (k_x,y) = [K_0 (|k_x y|) - K_0 (|k_x| \sqrt{4 t^2+y^2})] /2 \pi t$, where $K_0$ is the zeroth order modified Bessel function of the second kind. Using the uniform limit of $4 \pi t G_{zz} = \ln (1 + 4 t^2/y^2)$, we can calculate the static demagnetization field when the magnetization is uniformly magnetized along $\hat {y}$,
\begin{equation}
- \pi d_0 = \frac {w _-} {2} \ln \left(1 + w _- ^{-2}\right) + \frac {w _+} {2} \ln \left(1 + w _+ ^{-2}\right) - \tan ^{-1} w ^{-1} _- - \tan ^{-1} w ^{-1} _+
\label{yd0}
\end{equation}
with $w_\pm = (w \pm y)/ 2 t$. As swhon in Fig. \ref{ydemag}, $-d_0$ monotonically decreases from $y = 0$ to $y = \pm w$ and can act as a symmetric confining potential for spin wave. The discontinuity of $-d_0$ at $y = \pm w$ satisfies the boundary condition imposed by the continuity of the induction (\textbf {B}) field. Without this discontinuity, $d_0$ would be a continuous function over the whole $y$-space. Other components cannot be expressed explicitly, but their dependence on $k_x$ is easily seen as ($i, j = x, y$) $G_{ij}(k_x,y) = \int {d k_y} G_{ij} (\textbf{k}) \exp (i k_y y) /2 \pi$. From those expressions, it is obvious that the diagonal components are still real, but $G_{xy}$ is pure imaginary, $G^*_{xy} (k_x, y) = - G_{xy} (k_x, y)$, which means that the diagonal components are time-reversal even and the only non-zero off-diagonal component is odd under time-reversal.

Due to the fact that $f_k$ depends only on $k$, $G_{ij}$ are invariant under 2D inversion, $G_{ij} (k_x, y) = G_{ij} (-k_x, -y)$, i.e. the parity of $G_{ij}$ is $P = +1$. Separate parity $P_x$ or $P_y$ for inversion along $x$ or $y$ can also be defined with the relation $P = P_x P_y$: The diagonal components have parity $P_{x,y} = +1$, $G _{ii} (k_x, y) = G _{ii} (-k_x, y) = G _{ii} (k_x, -y)$ with $i = x$, $y$, or $z$, and the off-diagonal component has parity $P_{x,y} = -1$, $G _{xy} (k_x, y) = - G _{xy} (- k_x, y) = - G _{xy} (k_x, -y)$. This inversion symmetry is also obeyed in coordinate space \cite{Guslienko11}, $G_{zz} = [\rho^{-1} - \sqrt {4t^2+\rho^2}]/4 \pi t$ and $G_{ij} = - \partial ^2 \phi/ \partial x _i \partial x_j$ ($x_{i, j} = x, y$) where the scalar function \cite{Guslienko11} $\phi = [\xi - \sqrt {1 + \xi^2} + \ln (\xi^{-1} + \sqrt {1+\xi^{-2}})]/2\pi$ with $\xi = \rho/2t$. To get those expressions in coordinate space, it is easier to start from the expression for the Green function in 3D, $G_{ij} = -\partial^2 r^{-1}/\partial x_i \partial x_j$, and then average over the $z$ direction \cite{Guslienko11}. Actually, the inversion symmetry of the 2D Green function follows from the inversion symmetry of the 3D function, as the average over thickness conserves inversion symmetry in plane. $G _{ij} (\bm {\rho})$ are needed for consideration of spin wave in magnetic thin slabs.

\section{Theoretical framework}
\label{sw}
Magnetization dynamics in monolayer CrSBr is governed by the Landau-Lifshitz-Gilbert (LLG) equation \cite{Gilbert04} $\dot {\textbf{m} } = \textbf{h} \times \textbf{m} + \alpha \textbf{m} \times \dot {\textbf{m} }$ with the damping constant $\alpha$. A dot over \textbf{m} denotes the partial derivative with respect to time, $\dot {\textbf{m}} = \partial \textbf {m}/ \partial t$. The effective magnetic field normalized to $M_s$ is
\begin{equation}
\textbf{h} = \textbf {h}_{a} +  \hat {y} h _b m _y + \hat {z} h _c m _z + \nabla ^2 \textbf{m} - \textbf {d} .
\label{heff}
\end{equation}
$\textbf{h}_a$ is the externally applied field, $h _{b,c} = K_{b,c}/K_d$ are the uniaxial anisotropy fields along $b$ and $c$ directions, and $K_d = \mu_0 M_s^2/2$, where $\mu_0$ is the vacuum permeability, is the dipolar energy density constant. $\nabla$ is the gradient operator and $\hat {z}$ is the unit vector along the $z$ ($c$) direction, which is perpendicular to the $xy$ ($ab$) monolayer plane. Corresponding to the dimensionless form of field, angular frequency $\omega$ (or time $t$) is measured in units of $\gamma M_s$ (or $2 \pi/\gamma M_s$), where $\gamma$ is the gyromagnetic constant, and length in units of the exchange length $l _{ex} = \sqrt {A/K_d}$.

The equilibrium configuration of magnetization $\textbf{m}_0$ is determined by $\textbf{m} _0 \times \textbf{h} _0 = 0$, or equivalently the minimization of the magnetic interaction energy corresponding to the static effective field $\textbf{h} _0$, which consists of $\textbf{h}_a$, the static dipolar field $-\textbf{d}_0$ and other fields as defined in Eq. (\ref{heff}). In contrast to $\textbf{h}_a$, even for a uniformly magnetized state, $\textbf{d}_0$ is non-uniform. Once $\textbf{m}_0$ is given, it is convenient to change to a coordinate system in which $\textbf{m}_0$ is everywhere along the third-axis \cite{Dugaev05}, $\textbf{m} _0 = \hat {e}_3$. In this local coordinate system, spin wave can be considered as excitation over $\textbf{m}_0$, $\textbf{m} = \hat {e}_3 + \textbf{m}_1$, together with the field decomposition $\textbf{h} = h _0 \hat {e}_3 + \textbf{h} _1$ to the first order of $\textbf{m}_1 = \hat {e}_1 m_1 + \hat {e}_2 m_2$, where $\hat {e}_1$ and $\hat {e}_2$ are two unit vectors perpendicular to $\hat {e}_3$, satisfying the right-handed rule $\hat {e} _1 \times \hat {e}_2 = \hat {e} _3$. The equation of motion for spin wave in the local system is then $ \dot {\textbf{m}} _1 = (\tilde {\textbf{h}} - h _0 \textbf{m} _1) \times \hat{e} _3$. In terms of the complex spin-wave wave function $\psi = (\psi _+, \psi _-) ^T$, where $\psi _+ = m_1 + i m_2$ and $\psi_-$ is $\psi_+$'s complex conjugate, $\psi _- = \psi_+ ^* = m_1 - i m_2$, the dynamics of spin wave is determined by the Bogoliubov-de Gennes (BdG) equation
\begin{equation}
- i \sigma _z \dot {\psi} = H \psi
\label{bdg}
\end{equation}
with the Hamiltonian
\begin{equation}
H = H_0 - \sigma _x H_1 - \sigma _y G_{12} *,
\label{herham}
\end{equation}
where $H_0 = h_0 - \nabla^2 - (h_ + - G_+ *)/2$ and $H_1 = (h_ - - G_- *)/2$. $\sigma _x$, $\sigma_y$ and $\sigma _z$ are Pauli matrices, $h_\pm = h_1 \pm h_2$ with $h_1$ and $h_2$ being the anisotropy fields along $\hat{e}_1$ and $\hat{e}_2$ axes respectively, and similarly $G_\pm = G_{11} \pm G_{22}$. $G_\pm$ and $G_{12}$ describe the dynamic dipolar field when convoluted with $\psi$. Depending on the considered geometry, the convolution operation can reduce to multiplication in $\textbf{k}$-space for an infinite film, 1D convolution over $y$ for a waveguide, or 2D convolution over $\bm{\rho}$ for a slab.

$\sigma_z$ on the left hand side of Eq. (\ref{bdg}) reflects simply the symplectic structure of classical Hamilton dynamics, as the standard symplectic matrix $i \sigma _y$ is related to $- i \sigma_z$ by a unitary transformation $U = \exp {(i \theta)} \exp {(-i \theta \sigma _x)} \exp {(-i \theta \sigma _z)}$ with $\theta = \pi/4$, $-i \sigma _z = U^\dagger i \sigma _y U$. $U$ corresponds to the change of variables from $\psi_\pm/\sqrt{2}$ to $m_{1,2}$. $H$ in Eq. (\ref{herham}) is Hermitian. However, the solution of the BdG equation usually does not satisfy the condition $\psi_+ = \psi_-^*$, which is needed for the spin-wave interpretation. This difficulty can be resolved by the particle-hole symmetry \cite{Wang22} of $H$, $\sigma _x H ^* \sigma _x = H$, which ensures that the physical spin-wave wave function can be constructed from $\psi + \sigma _x \psi ^*$. Mathematically, the particle-hole symmetry of $H$ is a direct result of the realness of $H_0$, $H_1$ and $G_{ij}$, and it is not the usual time-reversal symmetry for bosonic field \cite{Lein19}. Physically, the particle-hole symmetry of Eq. (\ref{bdg}) corresponds to the well-known fact that spin wave is the collective motion of individual magnetic moments, which precess around their equilibrium directions; the direction of precession is determined by the gyromagenetic ratio $\gamma$, implying that changing the sign of $\omega$ cannot change the direction of precession of the underlying moments and, correspondingly, the chirality of the resultant spin wave.

Explicit expressions of $G_\pm$ and $h_\pm$ for field applied along three principal crystallographic axes are listed in Table \ref{ghpm}, where  $\textbf{m}_0$ is assumed to be parallel to $\hat {h}_a$ everywhere. This assumption of a uniform $\textbf{m}_0$ is valid for infinite films and waveguides, as $G_{xy} \propto k_x$ is zero when $k_x = 0$. For slabs, $G_{xy} \neq 0$ and its contribution to $\textbf {d}_0$ will introduce magnetization curling in the ground state, the effect of which on spin wave will be discussed briefly in Sec. \ref{swslab}. Table \ref{ghpm} shows clearly that $G_{12}$ is not zero only when $\hat {h}_a = \hat {z}$. Then, according to the symmetry of $G_{ij}$ discussed in Sec. \ref{mgf}, the Hamiltonian $H$ is invariant under inversion along $y$ when $\textbf{h}_a$ is applied in-plane and every magnon state should have a definite parity $P_y = \pm 1$, in addition to the 2D parity $P=\pm 1$. When $\hat {h}_a = \hat {z}$, $G_{xy}$ violates the inversion symmetry along $y$ generally, but it vanishes at $k_x = 0$. Hence we can still label magnon states in waveguides by their $y$-parity $P_y$ at $k_x = 0$. For brevity, we will still refer to the parity $P_y = \pm 1$ of a band, bearing in mind that the parity $P_y$ refers to the whole band for $\textbf{h}_a$ in-plane while it is only exact for the $k_x$-zero state in the band when $\hat {h} _a = \hat {z}$.
\begin{table}[htb]
\begin{tabular}{|c|c|c|c|c|}
\hline
$\hat {h}_a$ & $G_\pm$ & $G_{12}$ & $h_\pm$ & $h_0$\\
\hline
$\hat {z}$ & $G_{xx} \pm G_{yy}$ & $G_{xy}$ & $\pm h_b$ & $h_a + h_c - d_0$\\
\hline
$\hat {y}$ & $G_{zz} \pm G_{xx}$ & 0 & $h_c $ &$h_a + h_b - d_0$\\
\hline
$\hat {x}$ & $G_{yy} \pm G_{zz}$ & 0 & $h_b \pm h_c$ &$h_a - d_0$\\
\hline
\end{tabular}
\caption{$G_\pm$, $G_{12}$, $h_\pm$, and $h_0$ for $\hat{h}_a= \hat {x}$, $\hat {y}$, and $\hat {z}$. $-d_0$ is the static dipolar field, while $G_\pm$ and $G_{12}$ correspond to the dynamic part.}
\label{ghpm}
\end{table}

\section{Spin wave in momentum space}
\label{swk}
For an infinite film, spin wave should have the form $\psi \propto \exp [i (\textbf{k} \cdot \bm{\rho} + \omega t)]$, as there is no boundary existent. The eigenequation for $\omega$ is $\omega \sigma _z \psi_k =  H (\textbf{k}) \psi_k$. The Hamiltonian in \textbf{k}-space, $H(\textbf{k})$, is obtained from Eq. (\ref{herham}) by substituting the Laplacian operator $\nabla^2$ with $-k^2$. A rotation in spinor space with angle $\alpha_k$ determined by $\tan 2 \alpha_k = G_{12}/ H_1$ will transform $H(\textbf{k})$ into a simpler form $H' (\textbf{k}) = U H U^\dagger = H_0 - r_k \sigma _x$, with $r _k= (H_1 ^2 + G_{12}^2) ^{1/2}$ and $U = \exp (i \sigma_z \alpha _k)$. Eigenvalues and eigenfunctions of $H(\textbf{k})$, which would govern the dynamics of electrons through the corresponding Schr\"{o}dinger equation with the same $H(\textbf{k})$, can be obtained using the eigenfunctions $\eta_\pm$ of $\sigma_x$ with eigenvalues $\pm 1$, $\sigma_x \eta_\pm = \pm \eta _\pm$. Eigenvalues for $H(\textbf{k})$ are $\nu_\pm = H_ 0 \mp r_k$ and the corresponding eigenfunctions $\psi _\pm = U^ \dagger \eta_\pm$ with the conventional, or Euclidean, normalization $\psi _\pm ^\dagger \psi _\pm = 1$ using unit matrix as the metric for inner product.

However, spin-wave eigenvalue is not the eigenvalue of $H(\textbf{k})$: Due to the noncommutativity between $\sigma_z$ and $\sigma_{x,y}$, $H(\textbf{k})$ and $\sigma_z$ cannot have  simultaneous eigenfunctions. Luckily, as the rotation matrix $U$ still commutes with $\sigma_z$, the dispersion relation for spin wave is simple to obtain: $\sigma _z H' = \lambda _k V \sigma _z V^{-1}$, where  $\lambda _k = (H_0^2 - r_k^2) ^{1/2}$ is positive, can be diagonalized by a symmetric matrix $V = \exp (\sigma_x \beta _k)$ with the angle $\beta_k$ given by $\tanh 2 \beta _k = r _k/H _0$ which, although is real itself, is an imaginary rotation angle in spinor space. $\lambda_k$ here is just the geometric mean of the two eigenvalues of $H(\textbf{k})$, $\lambda (k)= \sqrt{\nu_+ \nu_-}$, which is a special feature in $\textbf {k}$-space. Due to its dependence on $k_{x,y}^2$ only, $\lambda_k$ is parity degenerate and states with definite parity can be constructed as  the linear combination $\psi \pm \bar {\psi}$, where $\bar \psi$ is the wave function after inversion. The only problem for this construction of states with definite parity is associated with the presence of $G_{12}$, which breaks the 1D inversion symmetries along $x$ and $y$ and hence the inversion symmetry of $H (\textbf {k})$, albeit $\lambda_k$ is still invariant under 1D inversions. Spin-wave eigenfrequency can be either positive or negative, $\omega_\pm = \pm \lambda _k$. Actually, the negative frequency branch of the spin-wave spectrum is related to the positive branch by the particle-hole symmetry, and the negative branch is needed to extract $m_1$ and $m_2$ as was aforementioned just below Eq (\ref{herham}). Spin-wave eigenfunction is $\psi_\pm = \exp (- i \sigma_z \alpha _k) \exp (\sigma _x \beta _k) \xi _\pm$ with the normalization $\psi _\pm^\dagger \sigma_z \psi_\pm = \pm 1$ using $\sigma_z$ as the metric, similar to the Minkowski metric in special relativity. $\xi_\pm$ are eigenvectors of $\sigma_z$, $\sigma_z \xi_\pm = \pm \xi_\pm$. Due to the non-positive-definite metric $\sigma _z$, the inner product of $\psi_i^\dagger$ and $\psi_i$ is not always positive. From the relation $V \sigma_z V^\dagger = \sigma_z$, we can see why we need to use $\sigma_z$ to normalize the eigenfunction. Another benefit of using the metric $\sigma_z$ is that the Hermiticity of $H(\textbf{k})$ ensures that the eigenvalue of $\sigma _z H(\textbf{k})$ is real \cite{Mostafazadeh02}.

An interesting feature of the dipolar interaction can be appreciated by writing out explicitly the Hamiltonian for $\hat {h} _a = \hat {z}$,
\begin{equation}
H (\textbf{k}) = H_0 + \frac {\sigma _x} {2} (h_b + f_k \cos 2 \varphi) + \frac {\sigma _y} {2} f_k \sin 2 \varphi
\label{hso}
\end{equation}
where $H_0 = h_0 + h_c - 1 + h _e k ^2 - (h_b - f_k)/2$ and $G_+(k) = f_k$, $G_-(k) = f_k \cos 2 \varphi$, and $2 G_{xy}(k) = f_k \sin 2 \varphi$ have been used. Angle $\varphi$ defines the propagation direction of spin wave, $\tan \varphi = k_y/k_x$. This Hamiltonian resembles that of an electron moving in a uniform electric potential $H_0$ and a magnetic field $\propto \hat {x} (h_b + f_k \cos 2 \varphi) + \hat {y} f_k \sin 2 \varphi$, which can be viewed as the effective field of an artificial Rashba spin-orbit interaction $\bm {\sigma} \cdot (\textbf{p} \times \hat {z})$ with effective momentum $\textbf{p} = \hat {y} (h_b + f _k \cos 2 \varphi) - \hat {x} f_k \sin 2 \varphi$. In the case of restored in-plane isotropy, $h_b = 0$, only the ubiquitous dipolar field contributes to the spin-orbit coupling for magnons. In the limit $k \rightarrow 0$, $\textbf{p} \rightarrow (\hat {y} \cos 2 \varphi - \hat {x} \sin 2 \varphi ) t k + \hat {y} h_b$, making the analogy almost perfect for the $\textbf{k}$-dependent part of $\textbf{p}$, except for the angle $2 \varphi$, which should be $\varphi$ for momentum \textbf{k}. For large $k$, the effective spin-orbit field saturates to $\hat {x} (h_b + \cos 2 \varphi) + \hat {y} \sin 2 \varphi$. As a consequence of the appearance of the artificial Rashba spin-orbit field, a minimum of $\omega$ at finite \textbf{k} will usually develop, similar to the corresponding shifted dispersion relation to electrons under the influence of authentic Rashba interaction \cite{Wang20jmmm,Wang20,Wang20stt}.

The origin of the angle $2 \varphi$ rather than the usual $\varphi$ can be traced back to the rotation of $\varphi$ in the spinor space, instead of the usual 3D coordinate space. This is obvious if the unitary rotation by angle $\varphi$ in the spinor space is performed explicitly as $U ^\dagger \sigma _x U = \sigma _x \cos 2 \varphi + \sigma_y \sin 2 \varphi$ with $U = \exp (i \varphi \sigma _z)$. This doubling in rotation angle is just a manifestation of the two-fold covering of SO(3) by SU(2).

\section{Spin wave in a waveguide}
\label{swwg}

\begin{figure}\centering
\begin{minipage}[c]{0.8\linewidth}
\includegraphics[width=\linewidth]{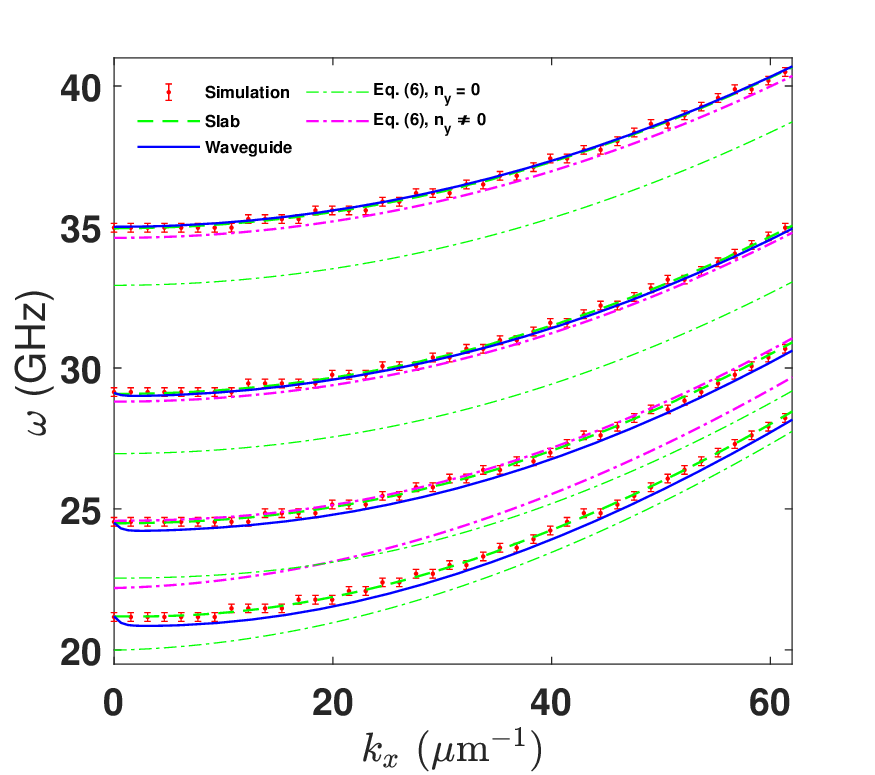}
\end{minipage}
\caption{$n ^ +$ spin-wave bands ($n = 0, 1, 2, 3$ from bottom to top) with field $\mu_0 H_a = 0.7$ T along $\hat{z}$ in a waveguide (solid lines) or a slab obtained from calculation (dashed lines) and micromagnetic simulation using OOMMF (symbols). Thick and thin dash-dotted lines correspond to Eq. (\ref{wz}) with $n_y \neq 0$ and $n_y = 0$, respectively. The error bar corresponds to the frequency resolution of the micromagnetic simulation.}
\label{zp}
\end{figure}

For a waveguide along $x$, spin wave should have the form $\psi \propto \mbox{exp}[i (\omega t + k_x x)]$, and the corresponding eigenequation for $\omega$ reads $ \omega \sigma _z  \psi = H (k_x) * \psi$, the analytical solution of which is difficult. We solved it numerically by expanding the wave function into a Fourier series and then diagonalizing the truncated secular matrix to obtain the discrete eigenvalue $\omega_i$ and wave function $\psi_i$. In the expansion of wave function, we used the quantized $w k_y = \pm n_y \pi, n_y = 0, 1, 2, \cdots$ for positive $y$-parity states in accordance with the free, or von Neumann, exchange boundary condition $\partial \psi/\partial y = 0$ at $y = \pm w$. Numerically, only one $n_y$ contributes dominantly to $\psi_i$ in $k_y$-space, and we will use the corresponding $n_y$ with a superscript denoting the $y$-parity to label the spin-wave frequency bands, unless otherwise stated. Only positive $\omega_i$ will be considered, as the negative $\omega_i$ are related to the positive $\omega_i$ through the particle-hole symmetry. The normalization of wave function is $< \psi _i| \sigma_z |\psi _j> = \pm \delta _{ij}$. For the fixed, or Dirichlet, boundary condition $\psi = 0$ at $y = \pm w$, the same set of $k_y$ can also be used for the expansion of wave function, but the resulted states' parity will be reversed, $n^\pm \rightarrow n^\mp$. The $n^\pm$ states with reversed parity for a specific $n$ are slightly different in frequency, although they are degenerate in frequency in an extended film. The monolayer thickness was set to $2t$ = 1 nm, which is slightly larger than the actual $c$ value, and the waveguide width was chosen to be $2w$ = 256 nm.

Figs. \ref{zp} shows the $n^+$ bands obtained by numerically diagonalizing $\sigma_z H$ with $\hat{h}_a = \hat{z}$. According to Table \ref{ghpm}, the dispersion relation for an extended film is explicitly
\begin{equation}
\omega^2 = (h_0 + k ^2 + G_{xx}) (h_0 - h_b + k ^2 + G_{yy} )
\label{wz}
\end{equation}
where $h_0 = h _a + h_c - 1$, corresponding to the static demagnetization factor $d_0 = n_z = G_{zz} (k=0) = 1$. For the film thickness of $2 t=$ 1 nm considered here, we had checked that the dispersion relation obtained by solving the boundary problem using linear superposition of eigenfunctions \cite{DeWames70,Hillebrands90} is numerically identical to Eq. (\ref{wz}) \cite{Supp}. As we had deliberately chosen $h_0 > h_b$ to ensure the stability of the ground state, the usual manifestation of the spin-wave spin-orbit coupling, $d \omega/dk_x <0$ which requires $h_b \cos ^2 \varphi > h_0$ for a fixed $\varphi \neq \pm \pi/2$ and $f_k \in [0, 1)$, cannot be observed for the dispersion given by Eq. (\ref{wz}). If the presence of boundary only introduces quantized wave vector as discussed in Ref. [\onlinecite{Guslienko02}], Eq. (\ref{wz}) should give a good description of the spin-wave bands, all of which should have positive group velocity. However, bands up to $2^+$ have negative group velocity around $k_x = 0$ and all the numerical bands have frequency higher than that predicted by Eq. (\ref{wz}). The downward turn of the lower frequency bands are caused by the dynamic dipolar field characterized by $G_{zz}$. The upward shift in frequency of the numerical bands in comparison to Eq. (\ref{wz}) is due to the static demagnetization factor \cite{Aharoni98} for a waveguide $n_z = 1 - n_y$ where $n_y = 8.76 \times 10 ^{-3}$ is actually smaller than the value $n_z = 1$ used in Eq. (\ref{wz}). Corresponding to the non-zero $n_y$, the modified $G_{zz} = n_z - f_k$ and $G_{yy} = n_y + f_k \sin ^2 \varphi$  have to be made to satisfy the constraint $(G_{zz} + G_{yy} + G_{xx}) (\textbf {k}) = 1$. A better description of the numerical result can be achieved with this modification: Except for the $0^+$ band, all higher $n^+$ bands can be approximated. The inability to describe the $0^+$ band by an effective $n_y$ can be attributed to the sensitivity of long-wavelength wave to the details at the boundary of the potential caused by the static dipolar field. If the wave length is small, the average potential will be determined mainly by the slowly-varying, or bulk, feature of the static potential, hence the description using $n_y$ preforms better for higher $n$. Although Eq. (\ref{wz}) with $n_y = 0$ agrees poorer with the numerical results, but it describes better the $n^-$ bands with the fixed boundary condition (not shown here). The reason behind this agreement is simply because that, due to the fixed boundary condition, spin wave is pushed away from the boundaries at $y = \pm w$, where the static dipolar field concentrates, as shown in Fig. \ref{ydemag}. Hence the effective potential felt by spin wave with fixed boundary condition is similar to that of an extended film, which can be described better by Eq. (\ref{wz}). The frequency splitting between bands with reversed parity, hence also corresponding to different boundary conditions (free or fixed), for specific $n$ can be roughly estimated between bands given in Fig. \ref{zp} with two choices for $n_y$. It is also interesting to note that, although in this case the potential caused by the dipolar field is locally parabolic, thus locally similar to the parabolic potential previously demonstrated to accommodate localized spin wave \cite{Tartakovskaya16}, there is no localized state observed in Fig. \ref{zp}, as the potential roughly scales as the inverse of $1-c y^2$ with constant $c <1$, which is not globally parabolic.

\begin{figure}\centering
\begin{minipage}[c]{0.8\linewidth}
\includegraphics[width=\linewidth]{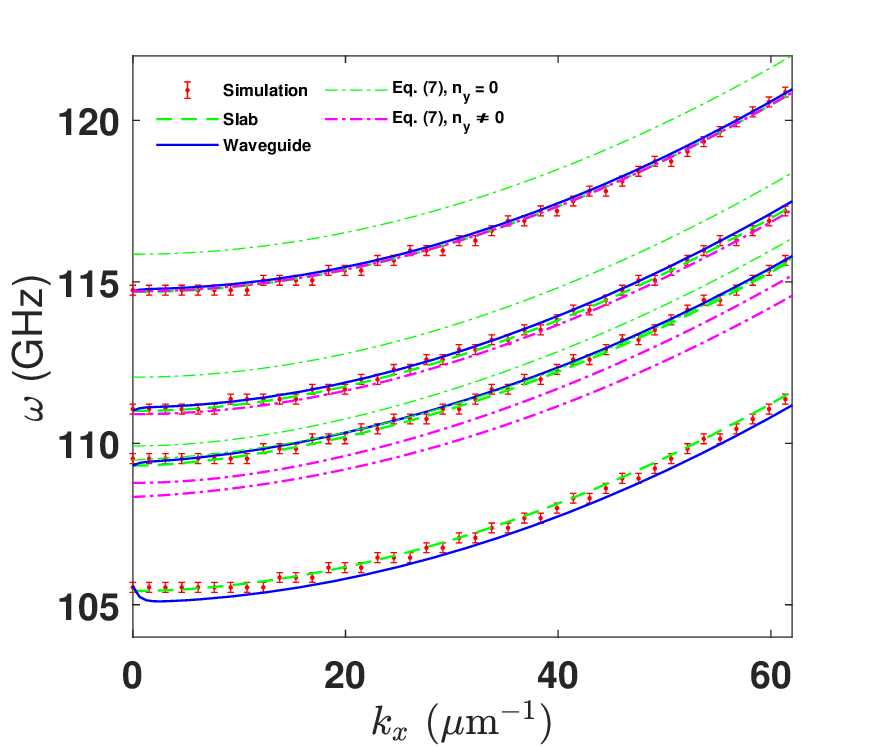}
\end{minipage}
\caption{$n ^ +$ spin-wave bands with field $\mu_0 H_a = 0.2$ T along $\hat{y}$ in the notation of Fig. \ref{zp}. Thick and thin dash-dotted lines correspond to Eq. (\ref{wy}) with $n_y \neq 0$ and $n_y = 0$, respectively. The lowest band is a localized band, while the other bands correspond to $n = 0, 2, 3$ from bottom to top}
\label{yp}
\end{figure}

When $\hat{h}_a = \hat{y}$, i.e. in the Damon-Eshbach geometry \cite{Eshbach60,Damon61}, the obtained energy bands are shown in Fig. \ref{yp}. In this case, the lowest band is the Damon-Eshbach mode, which is a localized mode (See Sec. \ref{deorigin}). The extended bands start from $0^+$, as usual, but with the $1^+$ band missing; So the band just above the second ($0^+$) band is the $2^+$ band. The Damon-Eshbach band is the only band that has a local minimum; Other bands are monotonic functions of $k_x$. Using Table \ref{ghpm}, the dispersion curves for an extended filem are given by
\begin{equation}
\omega^2 = (h_0 - h_c + k ^2 + G_{zz}) (h_0 + k ^2 + G_{xx}),
\label{wy}
\end{equation}
which are all above the numerical bands. When the dipolar field is considered using the $n_y$ given above, the static field is reduced to $h_0  = h _a + h_b -n_y$, which clearly shows that the inclusion of $n_y$ will reduce the frequency $\omega$ and can explain the reversal of order between the numerical results and Eq. (\ref{wy}) with $n_y =0$. But the lowering of the frequency by using an average $n_y$ is not enough to explain the numerical result; the lowest band is further below the lowest band predicted by Eq. (\ref{wy}) with $n_y \neq 0$, altough other bands, both calculated and simulated, agree fairly well with the $n_y \neq 0$ results. One additional feature for the Damon-Eshbach geometry is the reversal of order for the parity doublets with one $k_y$ value: The $n^+$ (free boundary) state is lower in frequncy than its $n^-$ (fixed boundary) partner, while it is the opposite case for $\hat {h}_0 = \hat {z}$. This is caused by the inverted, barrier shape of the static dipolar field, which is a potential well when $\hat {h}_0 = \hat {z}$: As the wave function of the $n ^+$ states does not vanish at the $y$-boundary due to the von Neumann boundary condition, the $n^+$ states tend to have a lower frequency than their $n^-$ partners whose wave function vanishes there.

The frequency bands for $\hat {h}_a = \hat {x}$ are given in Fig. \ref{xp}. There are two ($0^+, 1^+$) waveguide and one ($0^+$) slab bands showing negative group velocity in the vicinity of $k_x = 0$. In this configuration, we can easily see from the dispersion relation for an extended film,
\begin{equation}
\omega^2 = (h_0 - h_b + k ^2 + G_{yy}) (h_0 - h_c + k ^2 + G_{zz}),
\label{wx}
\end{equation}
that, with $\varphi = 0$, there is always an interval in $k_x$ where the group velocity is negative, as $h_0 > h_b$ is chosen in our calculation. In this case, as there is no static dipolar field for a waveguide, Eq. (\ref{wx}) with the modified dynamic $G_{yy}$ gives a very good approximation to the numerical result.

\section{Spin wave in a slab}
\label{swslab}
As an ideal, infinitely long, waveguide is difficult to realize experimentally, it is of interest to consider the spin-wave spectrum of a finite slab and to see how it is related to the spectrum of an ideal waveguide. To this aim, we performed theoretical calculation for a slab,  similar to the waveguide case shown in Sec. \ref{swwg}. We chose an aspect ratio of $l/w = 64$ with the same width $2w$ = 256 nm for the waveguide considered in Sec. \ref{swwg}. When $\hat {h}_a = \hat {z}$, due to the appearance of $G_{xy}$, $P_x$ and $P_y$ are not exact. But the effect of parity violation is small, and we can still seperate states according to their approximate $P_{x,y}$. As the frequency splitting of $P_x^\pm$ states is small, which is mainly caused by the smallness of $\pi/l$, we just averaged their values to get one frequency with definite $P_y$.

\begin{figure}\centering
\begin{minipage}[c]{0.8\linewidth}
\includegraphics[width=\linewidth]{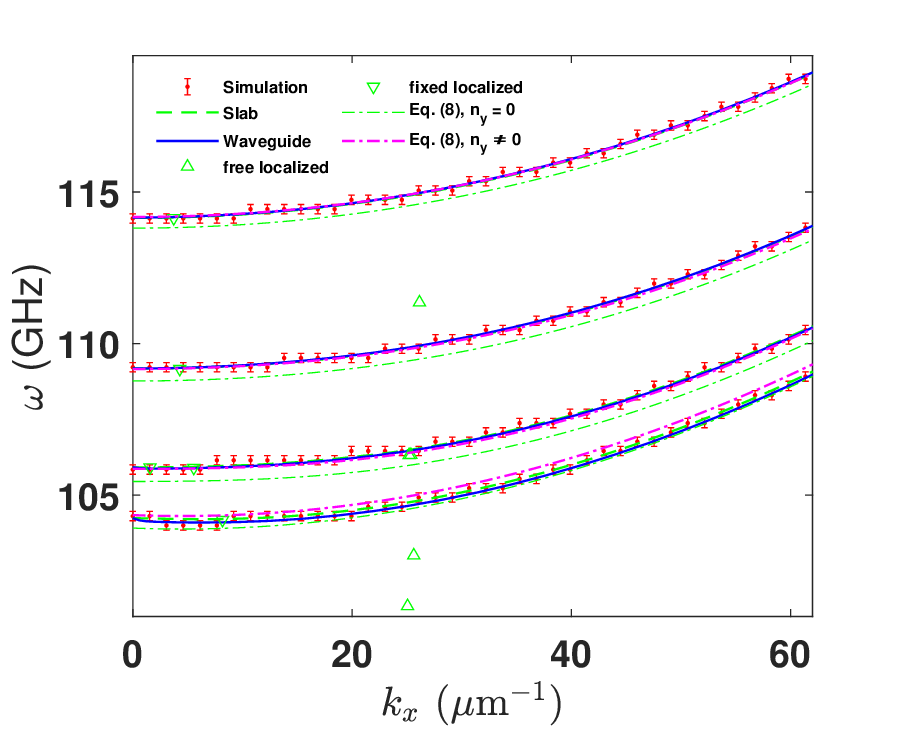}
\end{minipage}
\caption{$n ^ +$ spin-wave bands with field $\mu_0 H_a = 0.6$ T along $\hat{x}$ in the notation of Fig. \ref{zp}. Thick and thin dash-dotted lines correspond to Eq. (\ref{wx}) with $n_y \neq 0$ and $n_y = 0$, respectively. Localized states with positive $x$-parity for the slab geometry are shown as triangles.}
\label{xp}
\end{figure}

One benefit provided by the finite size of the slab is that the same spin-wave spectrum can be investigated by micromagnetic simulation, as an additional check of our theoretical calculation. We used OOMMF \cite{OOMMF} for this purpose. The simulated geometry is identical to our theoretical calculation, i.e. a monolayer CrSBr with length $2l= 16384 \Delta$ and width $2w = 256 \Delta$, where $\Delta = 1$ nm is the mesh size, which coincides with the monolayer thickness $2t = \Delta$. $\Delta$ used here is smaller than the exchange length, $l_{ex} \sim$ 4 nm, calculated using $K_d$. Comparison simulations were carried out to check that increase of $\Delta$ to 4 nm does not change the obtained dispersion relations substantially. A uniform state magnetized along one of three principal crystallographic axes is chosen as the initial state for subsequent time evolution over 1 ns with $\alpha$ set to 1. The obtained state is then minimized by conjugate gradient (CG) method as implemented in OOMMF. The CG minimization is just employed to ensure that the time-evolved state with finite $\alpha$ is still stable when $\alpha$ is set to zero for spin-wave simulation. Spin wave was excited by an oscillating field $ h_0 \mbox{sinc} (\omega_c t_r) \mbox{sinc} (k_x^c x_r) f _\pm (y)$ with $h_0 =  10^2$ A/m and $f_+ = \mbox {sinc} (k_y^c y_r)$ or $f_- = \mbox {sinc} (k_y^c y_r/2) \sin (k_y ^c y_r/2)$ for $P_y^\pm$ spin wave. The oscillating field is perpendicular to the applied field $\textbf{h}_a$: When $\hat {h}_a = \hat {z}$, it is parallel to $\hat {y}$; otherwise, it is parallel to $\hat{z}$. $x_r = x - x_0$, $y_r = y - y_0$ and $t_r = t - t_0$ are used to peak the excitation field at the centre of both time and space. The cut-off frequency is $\omega_c/2 \pi =$ 25 GHz, and the cut-off wave numbers are $k_x^c=$ 80 $\pi/l$ and $k_y^c=$ 12 $\pi/w$. A square-wave distribution in both frequency and $\textbf{k}$ was emulated by the sinc function, $\mbox {sinc} (x) = \sin(x)/x$, while $f_-$ was used to approximate an antisymmetric uniform distribution in $k_y$ up to $k_y^c$. Once the ground state was determined, the whole system was evolved in time with the oscillating field turned on for 20.48 ns with zero damping, outputting snapshots of magnetization state every 20 ps. Distribution of the spin-wave amplitude in $\omega$ and $k_x$ was obtained from fast Fourier transform of the output snapshots with respect to both $t$ and $x$, at $y = \pm (w - \Delta/2)$ ($\pm \Delta/2$) for the $P_y^+$ ($P_y^-$) modes.

Qualitative, the simulated ground state with $\hat {h} _a = \hat {x}$ or $\hat {y}$ is different from that with $\hat {h} _a = \hat {z}$. For the field $\mu_0 h_a = 0.7$ T applied along $\hat {z}$, the ground state is uniformly magnetized along $\hat {z}$, as in this case, the static dipolar field is simply antiparallel to $\textbf{h}_a$ due to the block-diagonal structure of $G_{ij}$. However, a uniform state magnetized in-plane along either $\hat {x}$ or $\hat {y}$ will become unstable, as the non-vanishing $G_{xy}$ will induce an additional  dipolar field component orthogonal to the uniform magnetization, and the ground state changes to the so-called 'flower state' \cite{Rave98} with curling around the four corners when $\hat{h}_a = \hat {x}$ or $\hat {y}$. From the study of spin wave in magnetization textures \cite{Dugaev05} such as skyrmions \cite{Wang22} and domain walls \cite{Wang17}, it is well known that magnetization textures will induce an emergent gauge field for magnons, further complicating the spin-wave dynamics. The correction induced by the emergent gauge field was neglected in our calculation of the dispersion relation simply because its contribution to the Hamiltonian matrix is negligible \cite{Supp}. As aforementioned, the parity-violation $G_{xy}$ is static and its effect on spin wave is mediated through the induced magnetization texture in the ground state, which is not symmetric under 1D inversion. But the non-uniformity is extremely weak since the applied in-plane field is large; The effect of $G_{xy}$ is mostly suppressed by $\textbf{h}_a$. When $\hat{h}_a = \hat{z}$, the situation is different: $G_{xy}$ acts as a dynamic dipolar field rather than being static and it is explicitly included in $H$.

The simulated bands are plotted together with the calculated bands in Figs. \ref {zp}, \ref{yp}, and \ref{xp}. They agree with each other within the simulation uncertainty. Except for the lowest one or two bands, the band structure for all three directions of \textbf {h}$_a$ can be better described by non-zero in-plane demagnetization factors, similar to Sec. \ref{swwg}, but with a slightly different \cite{Aharoni98} $n_y = 8.74 \times 10 ^{-3}$ and non-zero $n_x = 1.31 \times 10 ^{-4}$. The additional modification to the Green function corresponding to $n_x$ is $G_{xx} = n_x + f_k \cos ^2 \varphi$. As the difference in value between $n_y$ and $n_x$ for the slab geometry considered here and the waveguide geometry in Sec. \ref {swwg} is small and the resulted change is well within the plotted line thickness, we used the values give here even for the waveguide case in Sec. \ref {swwg}.

The overall agreement between slab and waveguide spin-wave bands is reasonable in Fig. \ref{zp}, especially for bands with $n > 0$. However, there are discrepancies at small $|k_x|$, where the $0^+$ waveguide band has a deeper and wider dip but the corresponding band given by simulation is rather shallow and narrow. Until the $2^+$ band, the waveguide bands still have a local minimum. This demonstrates clearly that the aspect ratio used in the simulation, $l/w=64$, is still not large enough to capture the physics for an infinitely-long waveguide. This feature is shared by the other two cases: There is only one ($0^+$) slab band showing negative group velocity in the vicinity of $k_x = 0$ for $\hat {h} _a = \hat {x}$, while  the group velocity of all slab bands is positive with $\hat {h} _a = \hat {y}$.

An outstanding feature of the slab geometry is the appearance of localized states in band gap of the calculated spectrum with $\hat {h} _a = \hat {x}$. They are localized about the two $x$-edges at $x = \pm l$. Those localized states are not observed in the simulated spectrum, possibly due to their localized character and a correspondingly small excitation efficiency in response to the applied field. A localized field can be used to excite them in simulation \cite{Supp}. Experimentally, those localized states may be observed by Brillouin light scattering \cite{Copus24}.

\begin{figure}\centering
\begin{minipage}[c]{0.8\linewidth}
\includegraphics[width=\linewidth]{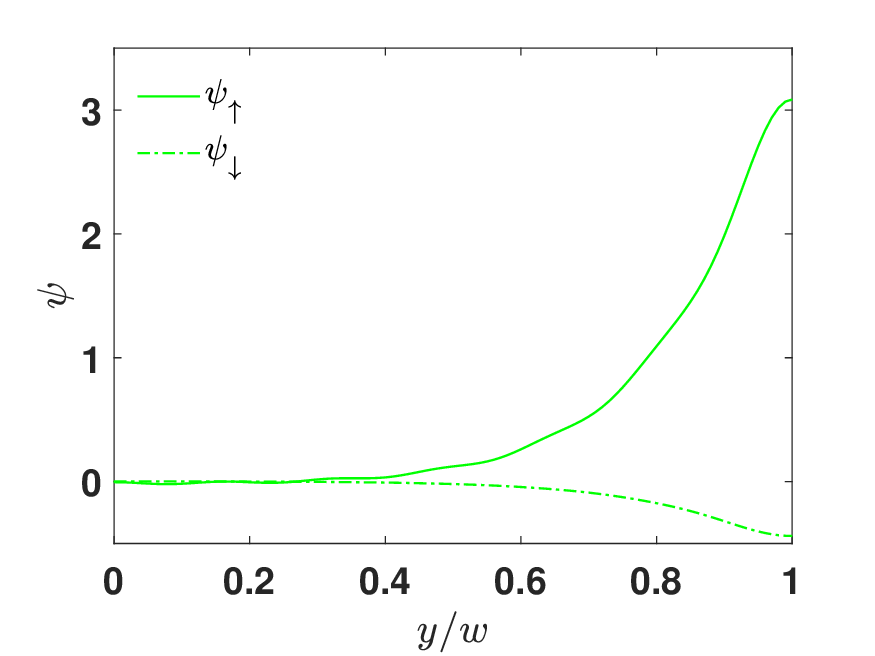}
\end{minipage}
\caption{Damon-Eshbach spin-wave wave function at $k_x = 0$.}
\label{depsi}
\end{figure}

\section{Origin of the Damon-Eshbach mode}
\label{deorigin}
The lowest band in Fig. \ref{yp}, which lies below the lowest dispersion curve for an infinite film, corresponds to the Damon-Eshbach mode \cite{Eshbach60,Damon61}, which is localized around the two edges at $y=\pm w$, as demonstrated by the spin-wave amplitude distribution shown in Fig. \ref{depsi}. Although it was discovered more than sixty years ago, its physical origin is still under debate. Inspired by its rather robust feature against back scattering, the authors of Ref. [\onlinecite{Yamamoto19}] suggested that the Damon-Eshbach mode is actually the edge mode determined by the topological winding number around four vortices that are time-reversal invariant in the reciprocal space. What we found is in stark contradiction with this topological origin: The existence of the Damon-Eshbach mode is simply caused by the confining potential provided by the static dipolar field, which acts as a potential in the BdG formulation of spin-wave equation of motion and is plotted in Fig. \ref{ydemag}. The barrier shape of the dipole potential indicates that the edges of a waveguide are more energetically favourable for spin wave to dwell on, just by reducing the total spin-wave energy in comparison to spreading around the centre of a waveguide, where the potential has a maximum.

In the usual bulk-boundary correspondence of topology, the appearance of topological edge modes localized around a topology defect also relies on the presence of interactions that interpolate continuously across the defect. However, according to Eq. (\ref{yd0}), the static dipolar potential is not continuous across the waveguide edges, as required by the continuity of the \textbf {B} field and shown in Fig. \ref{ydemag}. The discontinuity of the potential indicates that the Damon-Eshbach mode is not derived from the usual bulk-boundary correspondence of topology.

There are several possible reasons why the topological prediction does not realise in the case of exchange-dipole spin wave in a ferromagnetic waveguide. The most import one is the long-range characteristic of dipolar interaction, which makes the Hamiltonian a non-analytic function in both coordinate \cite{Supp} and reciprocal spaces. The aforementioned discontinuity in Fig. \ref{ydemag} is just a manifestation of this point in 1D coordinate space. The ambiguity in defining a continuous Hamiltonian in reciprocal space, as already pointed out in Ref. [\onlinecite{Yamamoto19}], is another manifestation. For consideration of topology, usually the Hamiltonian needs to be a continuous mapping between topological spaces \cite{Chiu16}. Secondly, the classification of topology \cite{Chiu16} according to the symmetry of the Hamiltonian for Fermions which obey anti-commutation relation is applied to spin wave \cite{Yamamoto19}, the quanta of which, magnons, should be Bosons and obey the commutation relation. Different statistics of Fermions and Bosons require different symmtry constraints for the single-particle Hamiltonian. Bosonic statistics will definitely modify the outcome of Ref. [\onlinecite{Yamamoto19}], which identified the topology of spin-wave single-particle Hamiltonian as class CI with time-reversal, particle-hole, and chiral symmetries \cite{Chiu16}. Even this classification of topology was controvertial: By considering the Krein structure of the spin-wave BdG equation, Ref. [\onlinecite{Lein19}] claimed that the topology class of spin wave is actually A with no symmetry at all \cite{Chiu16}, rather than CI. The classification of topology class A does not depend on the particle statistics, as it has no additional symmetry and hence no symmetry constraints need be considered. Last but not least, it is recently pointed out that bulk band topology does not necessarily lead to topologically protected  edge states with codimension 1 \cite{Xu25, Xu23}.

\section{Conclusion}
\label{conclusion}
Spin wave in monolayer CrSBr waveguides with field applied along all three principal crystalographic axes was investigated by a BdG formulation of the linearised LLG equation. Due to the spinor structure of the BdG equation and inversion symmetry, spin-wave eigenfrequencies form almost degenerate doublets with distinct parity, corresponding to different boundary conditions. Due to magnetostatic charge accumulation at the waveguide edges, static dipolar field acts as a confining potential for spin wave and localization of spin wave at the edges gives rise to the Damon-Eshbach mode. More localized states around the short edges of a ferromagnetic slab can be brought about by static dipolar field when the external field is applied along the long edges. Dynamic dipolar field couples the momentum and spin degrees of freedom of spin wave, resulting in spin-orbit coupling for magnons which is similar to that for electrons. The magnonic spin-orbit coupling  is usually manifested by a negative group velocity for spin wave with small $k_x$; spin wave inside this region of $k_x$ is commonly dubbed backward propagating.

\begin{acknowledgments}
The work was supported by the project Quantum Materials for Sustainable Technologies (QM4ST), no. CZ.02.01.01/00/22\_008/0004572, funded by the MEYS and co-funded by the EU.
\end{acknowledgments}

\newpage
\begin{center}Supplemental Material\end{center}
\renewcommand\theequation{S\arabic{equation}}
\renewcommand\thefigure{S\arabic{figure}}
\setcounter{equation}{0}
\setcounter{figure}{0}
\setcounter{section}{0}
\section{Spin wave dispersion from boundary condition}
\begin{figure}\centering
\begin{minipage}[c]{0.6\linewidth}
\includegraphics[width=\linewidth]{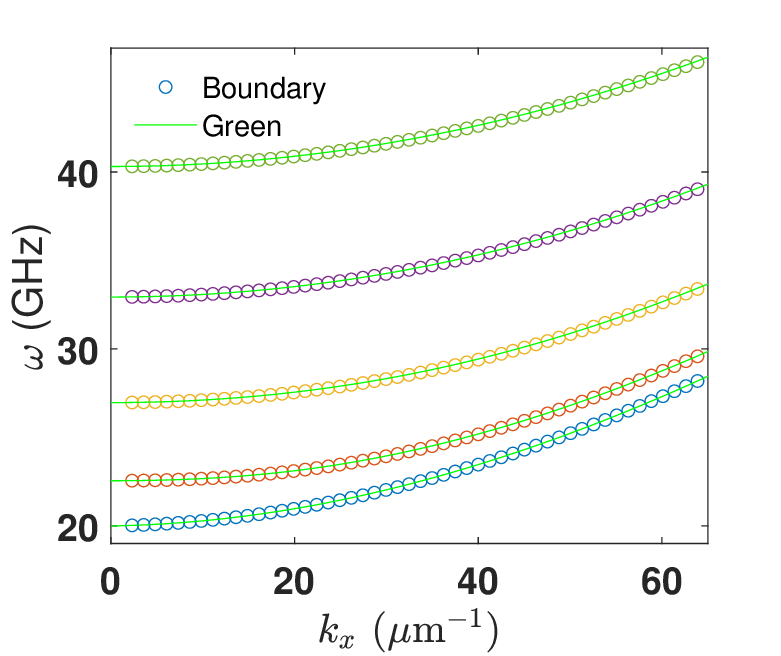}
\end{minipage}
\caption{Spin-wave dispersion curves obtained from boundary condition and Green function with $\hat {h}_a = \hat {z}$.}
\label{wktz}
\end{figure}

\begin{figure}\centering
\begin{minipage}[c]{0.6\linewidth}
\includegraphics[width=\linewidth]{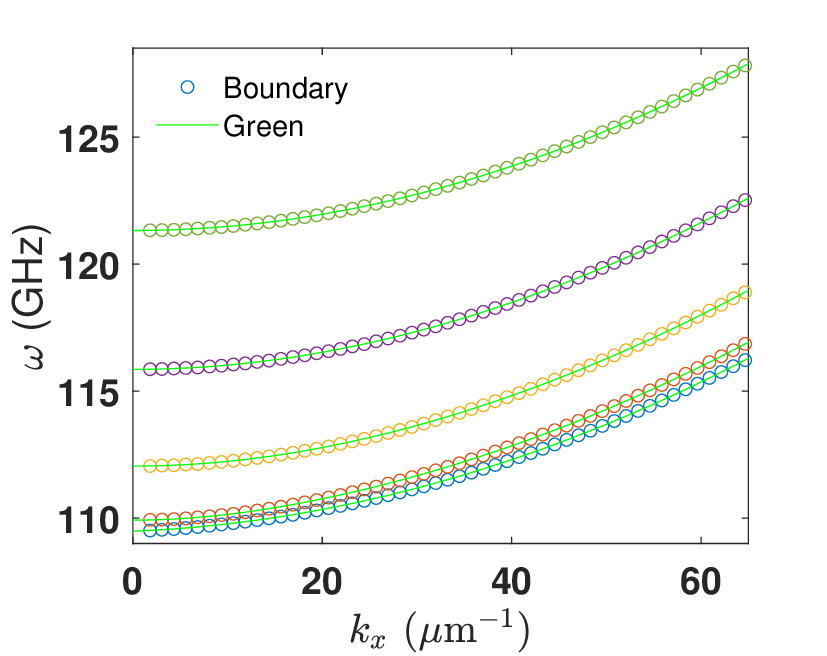}
\end{minipage}
\caption{Spin-wave dispersion curves obtained from boundary condition and Green function with $\hat {h}_a = \hat {y}$.}
\label{wkty}
\end{figure}

\begin{figure}\centering
\begin{minipage}[c]{0.6\linewidth}
\includegraphics[width=\linewidth]{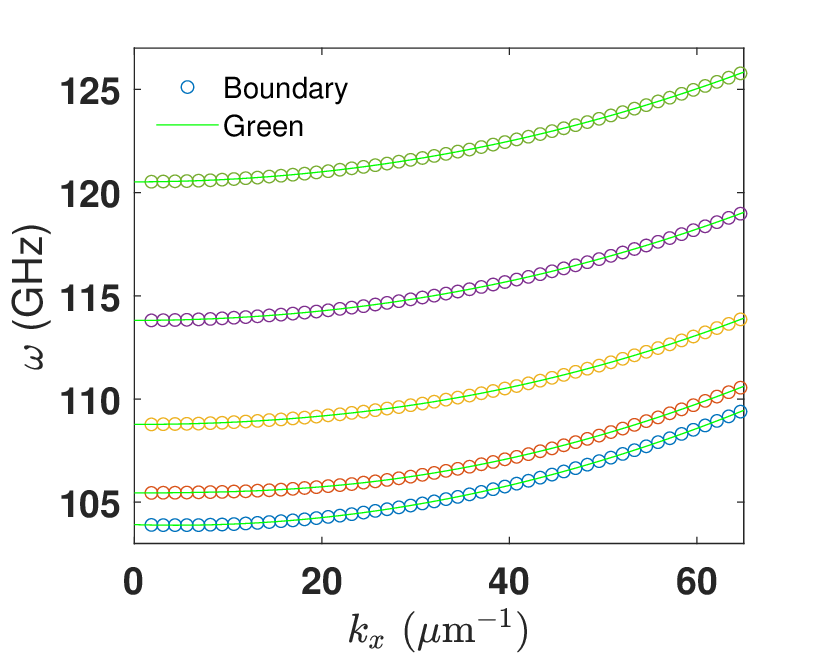}
\end{minipage}
\caption{Spin-wave dispersion curves obtained from boundary condition and Green function with $\hat {h}_a = \hat {x}$.}
\label{wktx}
\end{figure}

In the magneto-static approximation, the two magnetic Maxwell equations become $\nabla \cdot \textbf{B} = 0, \nabla \times \textbf{H} = 0$, as the mutual excitation between the magnetic and electric fields is neglected. With the definition of a scalar potential, $\textbf{H} = - \nabla \phi$, the equation for $\phi$ is $\nabla ^2 \phi = \nabla \cdot \textbf{m}$. For spin wave $\propto \exp {i ( {k} \cdot \textbf {x} + \omega t)}$, $k ^2 \phi = - i \textbf{k}  \cdot \textbf{m} _1$. $\textbf{h} =  - \textbf{k} \textbf{k}  \cdot \textbf{m} _1/k^2$. The singularity at $k^2 = 0$ is caused by the dispersion of the electromagnetic field in the magneto-static approximation, $k^2 = \omega ^2/c^2 = 0$, which is equivalent to $c \rightarrow \infty$. Effectively, the dynamic dipolar field can be expressed through a Green function $G_{ij} (k) = k_i k _j/k^2$, $i,j = 1, 2$. Using Table I and results of Sec. V in the main text, the spin-wave frequency is given by
\begin{equation}
\omega ^2 = (h_0 + k ^2 - h_1 + G_{11}) (h_0 + k ^2 - h_2 + G_{22}) - G_{12}^2.
\label{dispersion}
\end{equation}
Due to the cancellation enabled by the relation $G_{12}^2 = G_{11} G_{22}$, the equation for $k^2$ is of order three for fixed $\omega ^2$ and $k^2$ has three roots in the complex domain, one being real and the other two being complex conjugate to each other. We can label the $k_z$ corresponding to the three $k^2$ roots as $k_z ^i$, $i = 1$ to 6. Inside the infinite film ($z \in [-t, t]$), $\phi$, $m_1$ and $m_2$ can be expressed as the linear superposition of the six eigenmodes with different $k_z^i$,
$$\phi = c _i e ^{i k _i z} e ^{i (\omega t + k _y y)}, m _1 = a _i e ^{i k _i z} e ^{i (\omega t + k _y y)}, m _2 = b _i e ^{i k _i z} e ^{i (\omega t + k _y y)}.$$
Coefficients $a_i$ and $b_i$ for each mode $i$ are related to $c_i$ through the equation of motion for spin wave as
\begin{equation}
\frac {a _i} {c _i} = \frac {i (h_0 + k ^2 - h _2) k _1 - \omega k _2} {D}, \frac {b _i} {c _i} = \frac {\omega k _1 + i (h_0 + k ^2 - h_1) k _2} {D},
\end{equation}
where the denominator $D = \omega ^2 - (h_0 + k ^2 - h_1) (h_0 + k ^2 - h_2)$, the pole of which gives the spin-wave dispersion relation without considering the dynamic dipolar field. Above ($z \in (t, \infty)$) and below ($z \in (- \infty, -t)$), the scalar potential is given by $\phi = \phi _\pm e ^{\mp z k _\rho} e ^{i (\omega t + k _y y)}$. At the boundaries $z = \pm t$, the continuity of the normal component of \textbf {B} and the tangential component of \textbf {H} has to be satisfied, in addtion to the boundary condition for magnetization $\partial _z \textbf {m}_1 = 0$, which gives
\begin{eqnarray}
z = - t, \phi _- e ^{i k _2 z} = c _i e ^{i k _i z}, i k _2 \phi _- e ^{i k _2 z} = (i c _i k _i - b _i) e ^{i k _i z}, b _i k _i e ^{i k _i z} = 0, a _i k _i e ^{i k _i z} = 0.\nonumber\\
z = t, \phi _+ e ^{i k _1 z} = c _i e ^{i k _i z}, i k _1 \phi _+ e ^{i k _1 z} = (i c _i k _i - b _i) e ^{i k _i z}, b _i k _i e ^{i k _i z} = 0, a _i k _i e ^{i k _i z} = 0.
\end{eqnarray}
Setting the determinant of the secular matrix obtained from the coefficients of the linear system of equations gives the spin-wave dispersion relation. The results are shown in Figs. \ref{wktz}, \ref{wkty}, and \ref{wktx}. It can be seen that both approaches give almost identical results for monolayer CrSBr.

\section{Hamiltonian with in-plane magnetization curling}
\label{hgauge}
When $\hat{h}_a = \hat {x}$, the Hamiltonian components are
\begin{eqnarray}
H_0 = h_a \cos \theta + \frac{3} {2} h_b \sin ^2 \theta - \frac {h_b + h_c +  (\nabla \theta)^2} {2} - \nabla^2 - d_0 + \frac {G_{11} + G_{zz} *} {2},\nonumber\\
2 H_1 = h_c - h_b \cos ^2 \theta + (\nabla \theta)^2 + G_{11} - G _{zz} *,
\end{eqnarray}
where the static dipolar field $d_0 = [ (G_{xx}* \cos \theta) + (G_{xy} * \sin \theta)] \cos \theta + [(G_{yy} * \sin \theta) + (G_{xy} * \cos \theta)] \sin \theta$ and $G_{11} = \cos \theta (G_{yy} * \cos \theta - G _{xy} * \sin \theta) + \sin \theta (G_{xx} * \sin \theta - G _{xy} * \cos \theta)$ in the local coordinate frame where the local magnetization is along $\hat {e}_3$. Correspondingly, when $\hat{h}_a = \hat {y}$,
\begin{eqnarray}
H_0 = h_a \cos \theta + \frac{3} {2} h_b \cos ^2 \theta - \frac {h_b + h_c + (\nabla \theta)^2} {2} - \nabla^2 - d_0 + \frac {G_{zz} * + G_{22}} {2},\nonumber\\
2 H_1 = h_b \sin ^2 \theta - h_c - (\nabla \theta)^2 + G _{zz} * - G_{22}.
\end{eqnarray}
$d_0 = [(G_{yy} * \cos \theta) - (G_{xy} * \sin \theta)] \cos \theta + [(G_{xx} * \sin \theta) - (G_{xy} * \cos \theta)] \sin \theta$, and $G_{22} = \cos \theta (G_{xx} * \cos \theta + G _{xy} * \sin \theta) + \sin \theta (G_{yy} * \sin \theta + G _{xy} * \cos \theta)$. In both cases, the non-uniform distribution of magnetization is characterized by the angle $\theta$ which measures the in-plane rotation from $\textbf{h}_a$ to $\textbf{m}_0$. $\theta$ is of the order of 10$^{-2}$ or smaller for values of $h_a$ considered in the main text.

In comparison to the uniform case by setting $\theta$ to zero, we can see that the correction induced by in-plane magnetization curling is proportional to $\theta^2$. Using the texture profiles simulated from OOMMF, we computed the correction's distribution over the prism, excluding the dynamic dipolar field ($G_\pm$) contribution. It was found that the correction to both $H_0$ and $H_1$ is peaked around the four corners with magnitude of the order of at most 10$^{-3}$, which is what can be expected as $\theta$ is not zero only around the corners. Integration of the distribution over the prism region is used to estimate the correction to the Hamiltonian matrix. The correction is only of the order of 10$^{-7}$ for $H_0$, which is far below the resolution of frequency $3.3 \times 10^{-3}$ used in simulation. It is even one order smaller for $H_1$. Corrections with non-zero momentum transfer $\textbf{k}$ between the initial and final states are expected to have smaller magnitude due to the rapid oscillation of $\sin (\textbf{k} \cdot \bm{\rho})$ and $\cos (\textbf{k} \cdot \bm{\rho})$ in space. Dynamic dipolar field contribution to the correction should have comparable magnitude to the static one, if we notice that the dynamic correction is just the integral of the static one multiplied by a small number of the order of $\theta$.

\section{Negative parity bands}
\label{npbs}
\begin{figure}\centering
\begin{minipage}[c]{0.7\linewidth}
\includegraphics[width=\linewidth]{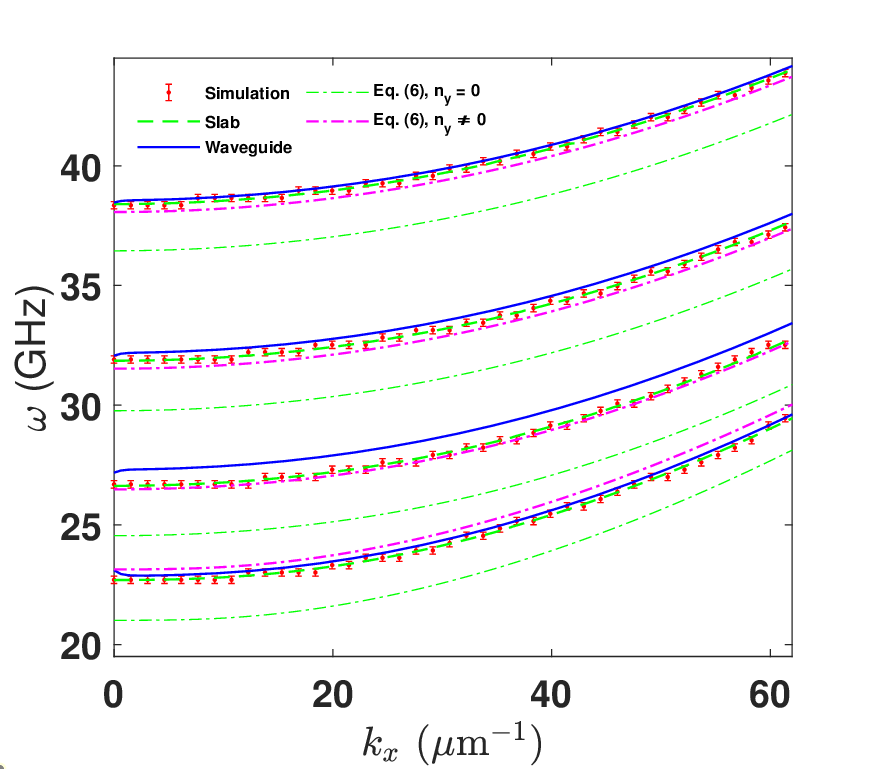}
\end{minipage}
\caption{$n ^ -$ bands for a monolayer CrSBr waveguide in the notation of Fig. 3.}
\label{zn}
\end{figure}

\begin{figure}\centering
\begin{minipage}[c]{0.7\linewidth}
\includegraphics[width=\linewidth]{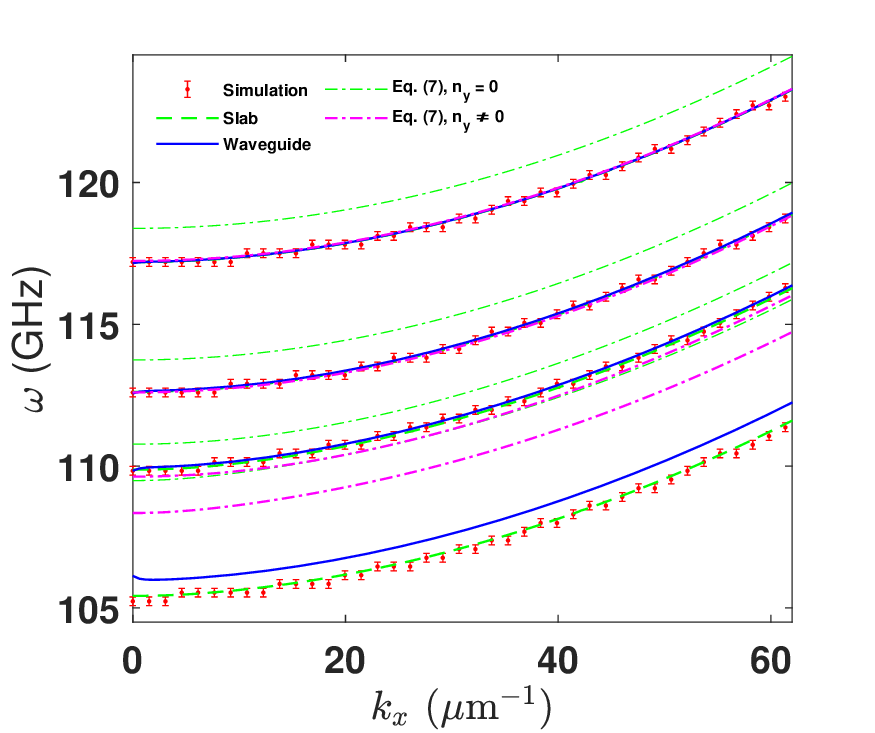}
\end{minipage}
\caption{$n ^ -$ bands for a monolayer CrSBr waveguide in the notation of Fig. 4.}
\label{yn}
\end{figure}

\begin{figure}\centering
\begin{minipage}[c]{0.7\linewidth}
\includegraphics[width=\linewidth]{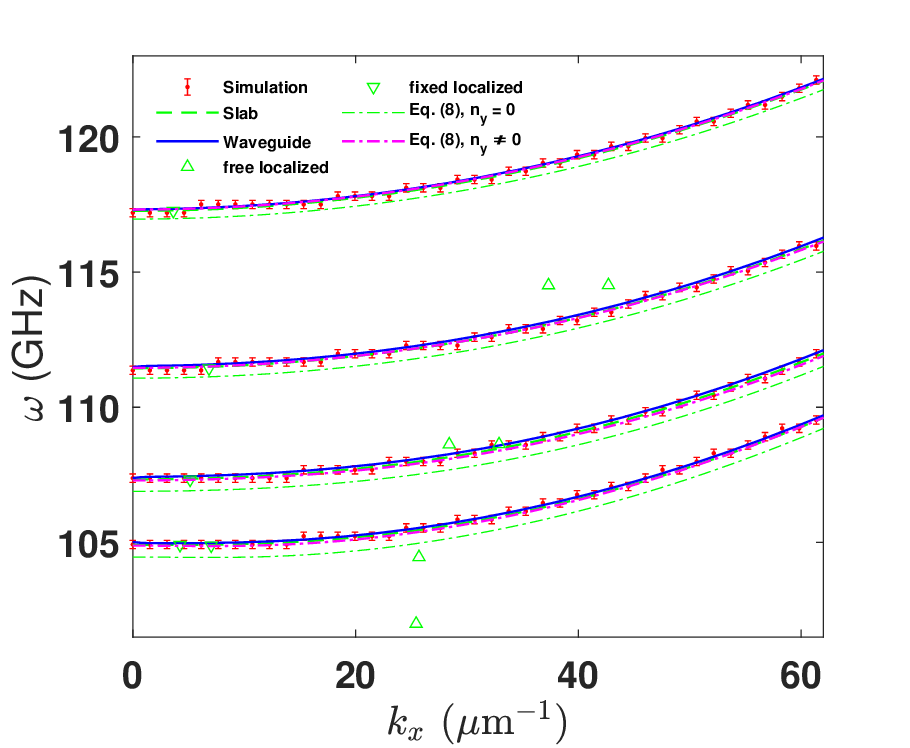}
\end{minipage}
\caption{$n ^ -$ bands for a monolayer CrSBr waveguide in the notation of Fig. 5.}
\label{xn}
\end{figure}

\begin{figure}\centering
\begin{minipage}[c]{0.7\linewidth}
\includegraphics[width=\linewidth]{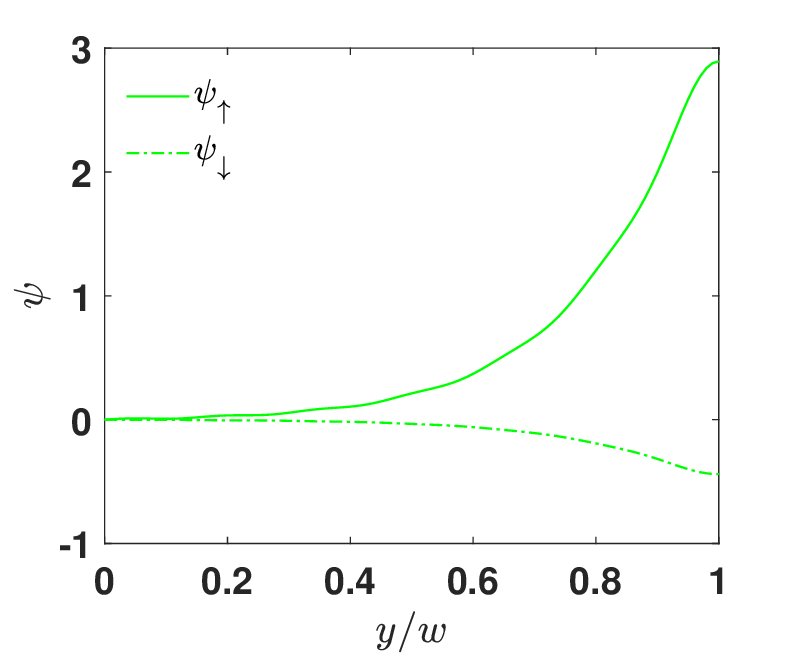}
\end{minipage}
\caption{Spin-wave wave function of the DE state with $k_x = 0$.}
\label{den}
\end{figure}

To obtain the $n_y^-$ bands, we need to use $w k_y = \pm (n_y - 1/2) \pi, n_y = 1, 2, \cdots$ for the expansion of wave function. For micromagnetic simulation, the $y$-profile of excitation field was changed to $f = \mbox {sinc} (k_y^c y_r/2) \sin (k_y ^c y_r/2)$ to approximate an antisymmetric uniform distribution in $k_y$ up to $k_y^c$. Other parameters remained the same as for the $n^+$ bands. Thus obtained frequency bands are shown in Figs. \ref{zn}, \ref{yn}, and \ref{xn}. The spin-wave distribution for the $n_y^-$ DE mode is given in Fig. \ref{den}

\section{Static demagnetization field}
\label{demag}
\begin{figure}\centering
\begin{minipage}[c]{0.6\linewidth}
\includegraphics[width=\linewidth]{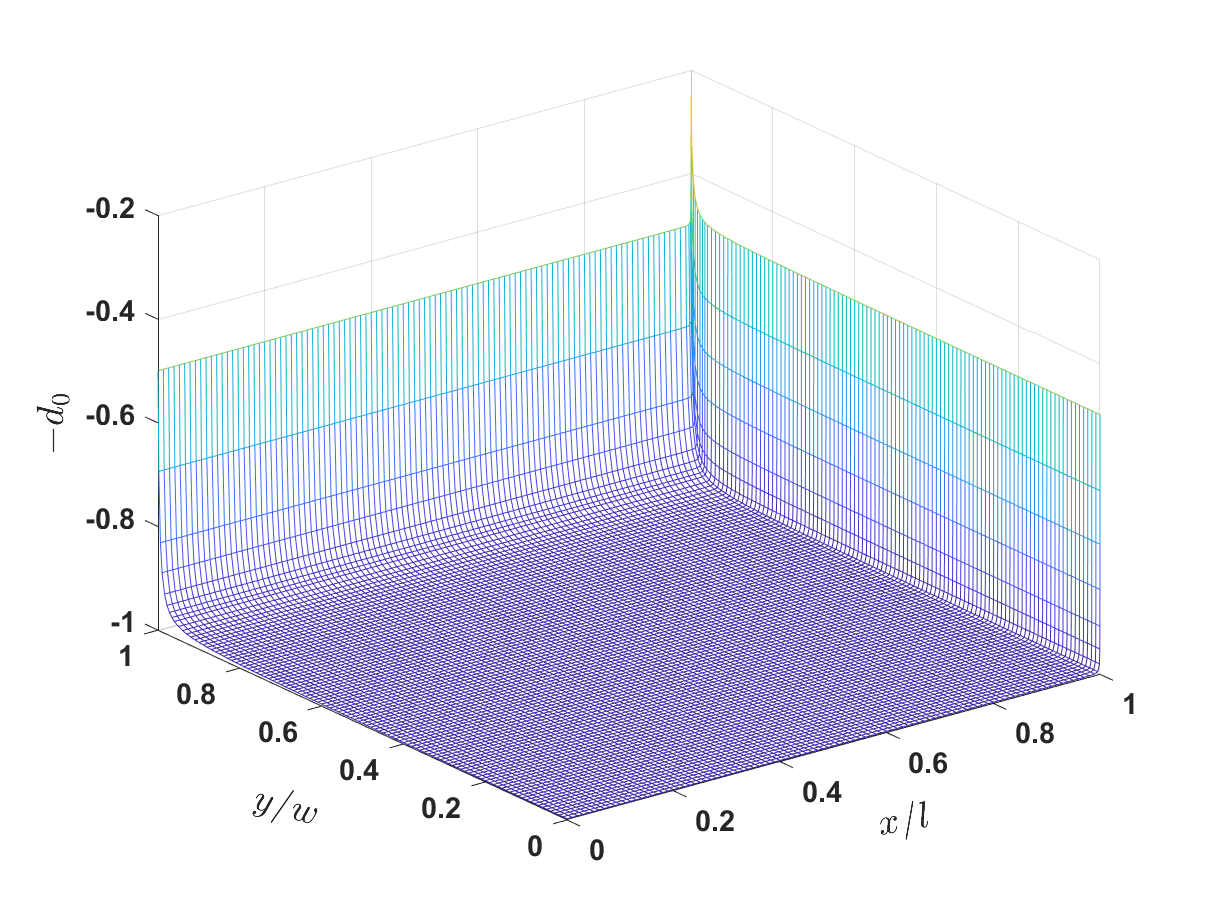}
\end{minipage}
\caption{Demagnetization field $- d_0^z$ for the slab dimension in the main text.}
\label{hdz}
\end{figure}

\begin{figure}\centering
\begin{minipage}[c]{0.6\linewidth}
\includegraphics[width=\linewidth]{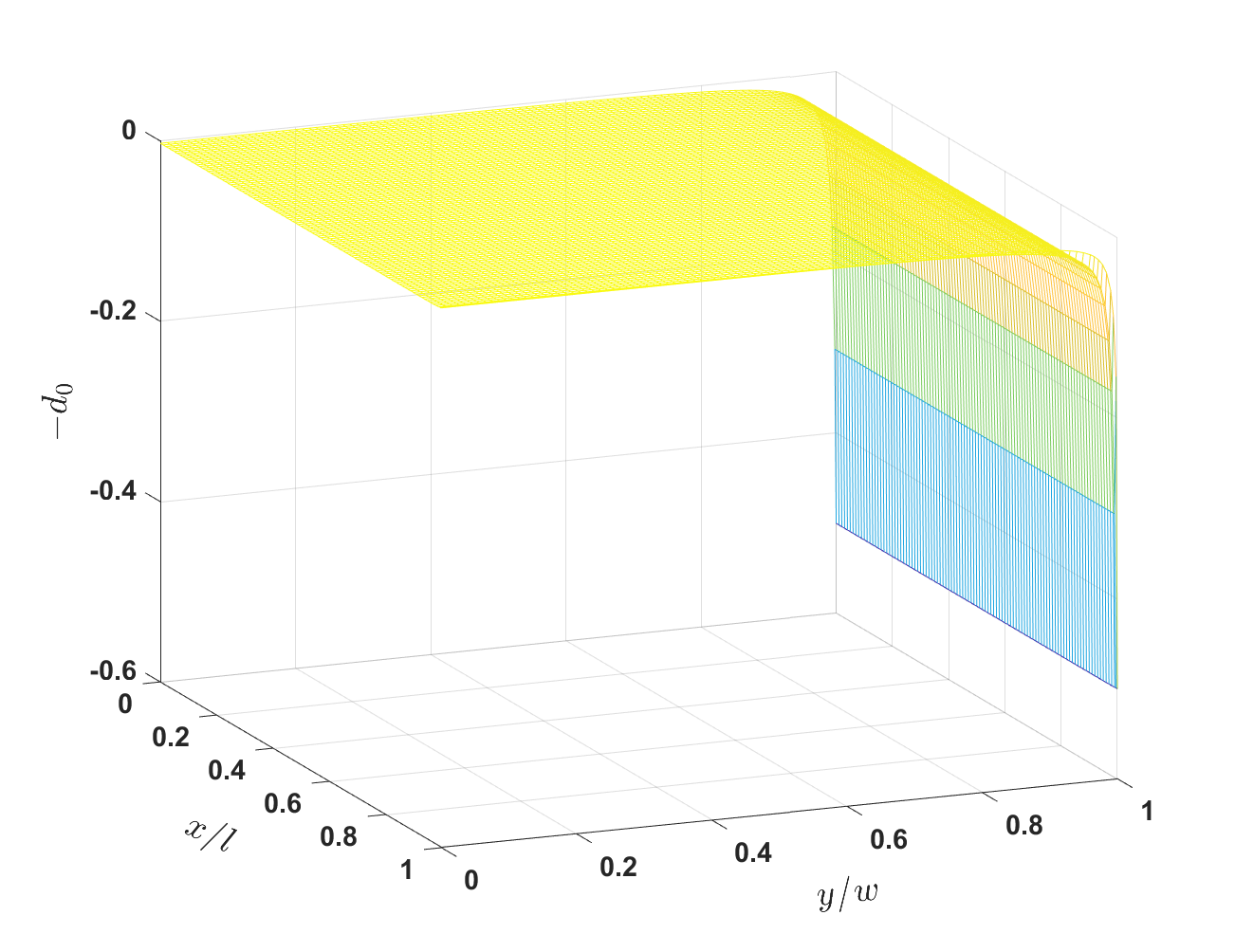}
\end{minipage}
\caption{Demagnetization field $- d_0^y$ for the slab dimension in the main text.}
\label{hdy}
\end{figure}

\begin{figure}\centering
\begin{minipage}[c]{0.6\linewidth}
\includegraphics[width=\linewidth]{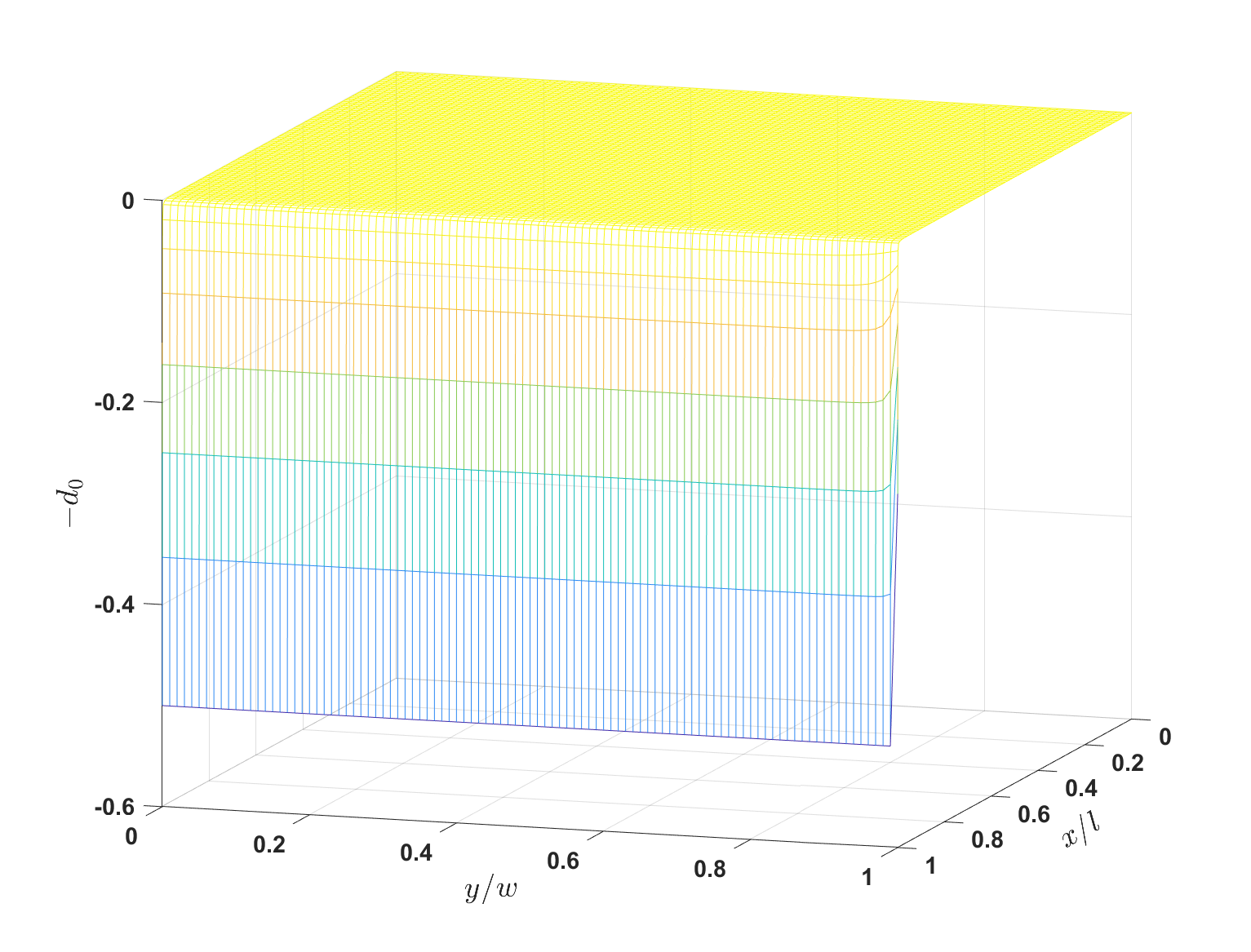}
\end{minipage}
\caption{Demagnetization field $- d_0^x$ for the slab dimension in the main text.}
\label{hdx}
\end{figure}

The static dipolar, or demagnetization, field for the slab dimension considered in the main text can be analytically calculated by integrating the Green function given therein over the slab region. The explicit expression for the integration of the $zz$ component is
\begin{eqnarray}
2 \pi d_0^z = w_+  \ln \frac{\sqrt{1+w_+^2+l_+^2}-l_+} {\sqrt{1+w_+^2+l_- ^2}+l_-} \frac{\sqrt{w_+^2+l_- ^2}+l_-} {\sqrt{w_+^2+l_+^2}-l_+} +  l_+  \ln \frac{\sqrt{w_+^2+l_+ ^2}+w_+} {\sqrt{1+w_+^2+l_+ ^2}+w_+}\nonumber\\
 +  l_-  \ln \frac{\sqrt{w_+^2+l_- ^2}+w_+} {\sqrt{1+w_+^2+l_- ^2}+w_+} +\tan ^{-1}  \frac{l_- w_+} { \sqrt{1+w_+^2+l_- ^2}} + \tan ^{-1}  \frac{l_+ w_+} { \sqrt{1+w_+^2+l_+ ^2}}\nonumber\\
 + (y \rightarrow -y) + l_+ \ln (1+l_+^{-2}) + l_- \ln (1+l_-^{-2}),
\end{eqnarray}
where $l_\pm = (l \pm x) /2 t$. For the $yy$ component, the integration gives
\begin{eqnarray}
2 \pi d_0^y = w_+  \ln \frac{\sqrt{1+w_+^2+l_+^2}+l_+} {\sqrt{1+w_+^2+l_- ^2}-l_-} \frac{\sqrt{w_+^2+l_- ^2}-l_-} {\sqrt{w_+^2+l_+^2}+l_+} - \tan^{-1}  \frac {2 w_+} {1-w_+^2}\nonumber\\
+\tan ^{-1}  \frac{l_-  \sqrt{1+w_+^2+l_- ^2}+l_-^2 + w_+^2} {w_+} + \tan ^{-1} \frac{l_+ \sqrt{1+w_+^2+l_+^2}+l_+^2 + w_+^2} {w_+} + (y \rightarrow -y).
\end{eqnarray}
$-d_0^z$ and $-d_0^y$ corresponds to the static demagnetization field acting on the magnetization when it is uniformly magentized along $\hat {z}$ and $\hat {y}$, respectively. The demagnetization field for magnetization along $\hat{x}$ can be obtained using the constraint $d_0 ^x + d _0 ^y + d_0 ^z = 1$. They are symmetric under inversion along both $x$ and $y$. Due to the long-range character of the dipolar field, the slope of $d_0^z$ at the four equivalent corners $x = \pm l$ and $y = \pm w$ is infinite, the value of $d_0^y$ depends on the path taken to approach the corners, and $d_0^x$ has both singular features. Their distribution is plotted in Figs. \ref{hdz}, \ref{hdy}, and \ref{hdx}. Integration of the off-diagonal element $G_{xy}$ gives the antisymmetric demagnetization field,
\begin{eqnarray}
2 \pi d_0^{xy} = \sqrt {1 + l_+^2 + w_+^2} - \sqrt {1 + l_-^2 + w_+^2} + \sqrt {l_-^2 + w_+^2} - \sqrt {l_+^2 + w_+^2}\nonumber\\
+ \ln \sqrt {\frac {w_+^2 + l_+^2} {w_+^2 + l_-^2}}  \frac {\sqrt {1 + w_+^2 + l_-^2} + 1} {\sqrt {1 + w_+^2 + l_+^2} + 1} - (y \rightarrow -y).
\end{eqnarray}
$d_0^{xy}$ is divergent at the corners of the slab. Due to the presence of $d_0^{xy}$, a uniform state magnetized in-plane is unstable against magnetization curling.

\section{Localized states}
\label{local}
\begin{figure}\centering
\begin{minipage}[c]{0.7\linewidth}
\includegraphics[width=\linewidth]{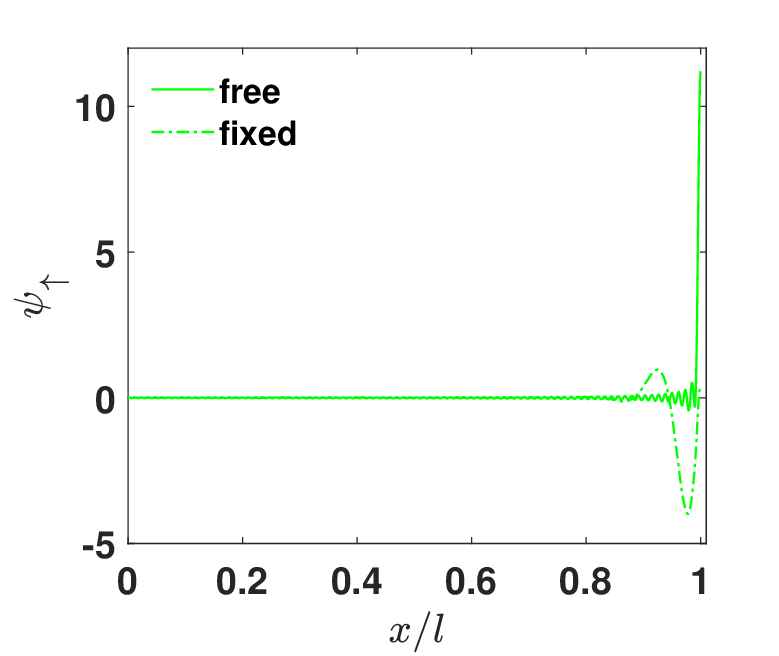}
\end{minipage}
\caption{Spin-wave amplitude at $y=0$ of the lowest free and fixed $n ^ +$ states in Fig. 5.}
\label{loc}
\end{figure}

\begin{figure}\centering
\begin{minipage}[c]{0.7\linewidth}
\includegraphics[width=\linewidth]{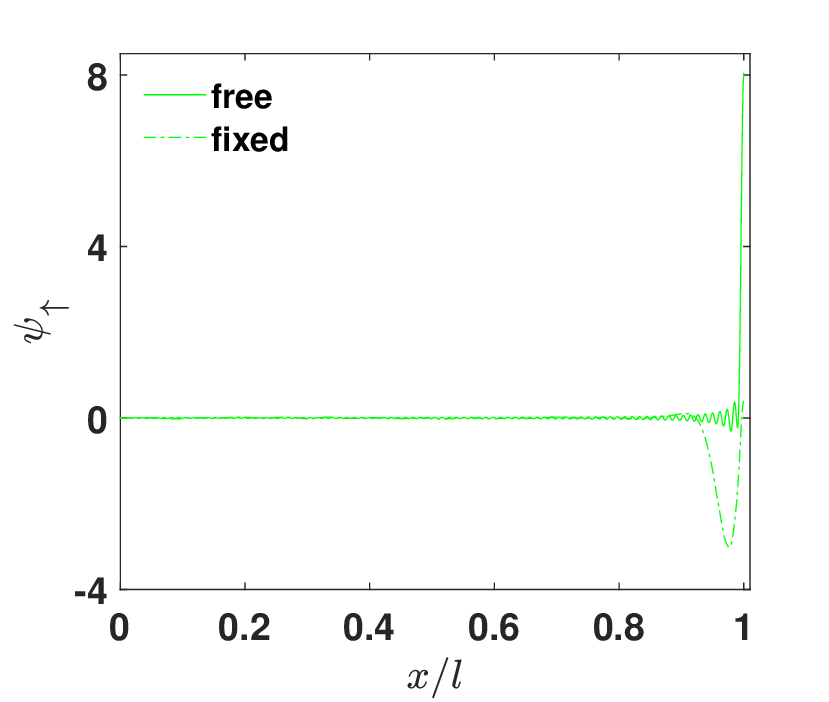}
\end{minipage}
\caption{Spin-wave amplitude at $y=w$ of the lowest free and fixed $n ^ -$ states in Fig. \ref{xn}.}
\label{lochp}
\end{figure}

Theoretically calculated spin-wave distribution for the lowest fixed and free localized states when $\hat {h}_a = \hat {x}$ is shown in Figs. \ref{loc} and \ref{lochp}.

\begin{figure}\centering
\begin{minipage}[c]{0.6\linewidth}
\includegraphics[width=\linewidth]{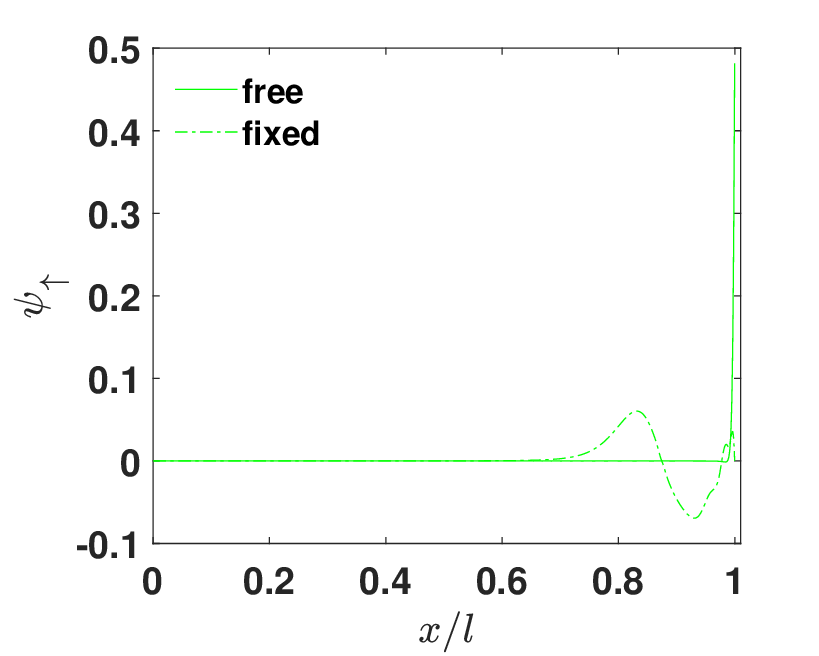}
\end{minipage}
\caption{$P_y^+$ and $P_x^+$ localized states from micromagnetic simulation at $y=\pm \Delta/2$. The frequency of the free (fixed) localized state is 16.0 (16.7) GHz.}
\label{locp}
\end{figure}

\begin{figure}\centering
\begin{minipage}[c]{0.6\linewidth}
\includegraphics[width=\linewidth]{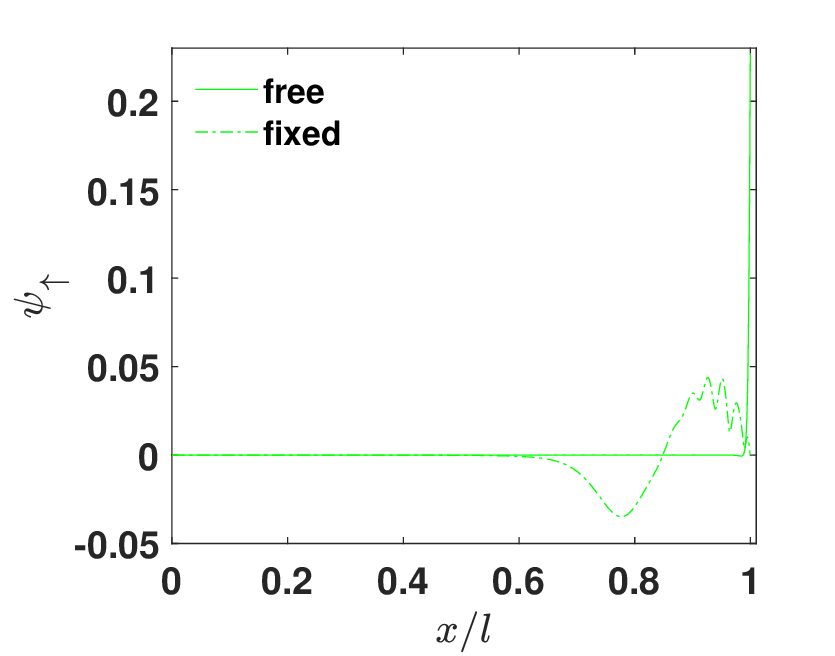}
\end{minipage}
\caption{$P_y^-$ and $P_x^+$ localized states from micromagnetic simulation at $y=\pm (w - \Delta/2)$. The frequency of the free (fixed) localized state is 16.1 (17.1) GHz.}
\label{locn}
\end{figure}

To observe the localized states in simulation, an oscillating field with frequency $f$ localized in one mesh cell at the boundary at $x = \pm l$ was applied, and the corresponding Fourier transform at the frequencies of the lowest localized states was obtained to extract the spin-wave amplitude. For $n_y^+$ states, the excitation field is uniform in $y$, while it is antisymmetric in $y$ with the sinusoidal profile $\sin \pi y_r/2$ around $y=0$. The obtained spin-wave distribution, after factoring out a high-$k_x$ component, is shown in Figs. \ref{locp} and \ref{locn}. The states correspond to the free boundary condition agree well with the theoretical calculation, but the agreement between states with fixed boundary condition is only qualitatively, especially for the $n_y^-$ state in Fig. \ref{locn}. One probable reason is the close vicinity of the localized states' frequency to the propagating states and to each other, as shown in Fig. \ref{xn}.

\end{document}